\documentclass[iop,onecolumn]{emulateapj}
\shorttitle{EFFECTIVE MODELS FOR GRAVITATIONAL LENSING}
\shortauthors{A. LAPI ET AL.}

\begin{document}
\title{Effective Models for Statistical Studies of Galaxy-Scale Gravitational Lensing}
\author{A. Lapi\altaffilmark{1,2}, M. Negrello\altaffilmark{3}, J.
Gonz\'alez-Nuevo\altaffilmark{4,2}, Z.-Y. Cai\altaffilmark{2}, G. De
Zotti\altaffilmark{3,2}, L.
Danese\altaffilmark{2}} \altaffiltext{1}{Dip. Fisica, Univ. `Tor Vergata',
Via Ricerca Scientifica 1, 00133 Roma, Italy} \altaffiltext{2}{Astrophysics
Sector, SISSA, Via Bonomea 265, 34136 Trieste,
Italy}\altaffiltext{3}{INAF-Osservatorio Astronomico di Padova, Vicolo
dell'Osservatorio 5, 35122 Padova, Italy} \altaffiltext{4}{Inst. de Fisica de
Cantabria (CSIC-UC), Avda. los Castros s/n, 39005 Santander,
Spain}

\begin{abstract}
We have worked out simple analytical formulae that accurately approximate the
relationship between the position of the source with respect to the lens
center and the amplification of the images, hence the lens cross section, for
realistic lens profiles. We find that, for essentially the full range of
parameters either observationally determined or yielded by numerical
simulations, the combination of dark matter and star distribution can be very
well described, for lens radii relevant to strong lensing, by a simple
power-law whose slope is very weakly dependent on the parameters
characterizing the global matter surface density profile and close to
isothermal in agreement with direct estimates for individual lens galaxies.
Our simple treatment allows an easy insight into the role of the different
ingredients that determine the lens cross section and the distribution of
gravitational amplifications. They also ease the reconstruction of the lens
mass distribution from the observed images and, vice-versa, allow a fast
application of ray-tracing techniques to model the effect of lensing on a
variety of source structures. The maximum amplification depends primarily on
the source size. Amplifications larger than $\approx 20$ are indicative of
compact source sizes at high-$z$, in agreement with expectations if galaxies
formed most of their stars during the dissipative collapse of cold gas. Our
formalism has allowed us to reproduce the counts of strongly lensed galaxies
found in the H-ATLAS SDP field. While our analysis is focussed on spherical
lenses, we also discuss the effect of ellipticity and the case of late-type
lenses (showing why they are much less common, even though late-type galaxies
are more numerous). Furthermore we discuss the effect of a cluster halo
surrounding the early-type lens and of a supermassive black hole at its
center.
\end{abstract}


\section{Introduction}\label{sect:intro}

\setcounter{footnote}{0}

The discovery rate of strong galaxy-scale lens systems has increased
dramatically in recent years mostly thanks to spectroscopic lens searches
and, most recently, to surveys of sub-millimeter galaxies. Spectroscopic
searches, such as the Sloan Lens Advanced Camera for Surveys (SLACS) survey
(Bolton et al. 2006, 2008; Auger et al. 2009) or the BOSS (Baryon Oscillation
Spectroscopic Survey) Emission-Line Lens Survey (BELLS; Brownstein et al.
2012) or the 'optimal line-of-sight' (OLS) lens survey (Willis et al. 2006)
or the Sloan WFC (Wide Field Camera) Edge-on Late-type Lens Survey (SWELLS;
Treu et al. 2011), rely on the detection of multiple background emission
lines in the residual spectra found after subtracting best-fit galaxy
templates to the foreground-galaxy spectrum.

Sub-millimeter surveys were predicted (Blain 1996; Perrotta et al. 2002,
2003; Negrello et al. 2007), and demonstrated (Negrello et al. 2010), to be
an especially effective route to efficiently detect strongly lensed galaxies
at high redshift because the extreme steepness of number counts of unlensed
high-$z$ galaxies implies a strong magnification bias so that they are easily
exceeded by those of strongly lensed galaxies at the bright end. Also,
gravitational lensing effects are more pronounced for more distant sources.
But high-$z$ galaxies are frequently in a dust-enshrouded active star
formation phase and therefore are more easily detected at far-IR/sub-mm
wavelengths, while they are very optically faint. Negrello et al. (2007)
predicted that about 50\% of galaxies with $500\,\mu$m flux densities above
$\approx 100\,$mJy would be strongly lensed, with the remainder easily
identifiable as local galaxies or as radio-loud Active Galactic Nuclei
(AGNs). This prediction was supported by the mm-wave South Pole Telescope
(SPT) counts (Vieira et al. 2010). But a spectacular confirmation came with
the results of the \textsl{Herschel} Astrophysical Terahertz Large Area
Survey\footnote{http://www.h-atlas.org/} (H-ATLAS; Eales et al. 2010) for the
Science Demonstration Phase (SDP) field covering about $14.4$ deg$^2$. Five
out of the 10 extragalactic sources with $S_{500\,\mu\rm m}\ga 100\,$mJy were
found to be strongly lensed high-$z$ galaxies, four are $z<0.1$ spiral
galaxies and one is a flat-spectrum radio quasar (Negrello et al. 2010).
Gonzalez-Nuevo et al. (2012) presented a simple method, the
\textsl{Herschel}-ATLAS Lensed Objects Selection (HALOS), aimed at
identifying fainter strongly lensed galaxies. This method gives the prospect
of reaching a surface density of $\sim 2$ deg$^{-2}$ for strongly lensed
candidates, i.e., the detection of $\sim 1000$ high-$z$ strongly lensed
galaxies over the full H-ATLAS survey area ($\approx 550$ deg$^{2}$; Eales et
al. 2010).

It should be noted that the (sub-)mm selected lensed galaxies are very faint
in the optical, while most foreground lenses are passive ellipticals (Auger
et al. 2009; Negrello et al. 2010), essentially invisible at sub-mm
wavelengths. This means that the foreground lens is `transparent' at (sub-)mm
wavelengths, i.e. does not confuse the images of the background source.
Therefore, the (sub-)mm selection shares with spectroscopic searches the
capability of detecting lensing events with small impact parameters and has
the advantage that, in most cases, there is no need to subtract the lens
contribution to recover the source images within the effective radii of the
lenses. Also, compared to the optical selection, the (sub-)mm selection
allows us to probe earlier phases of galaxy evolution which have typically
higher lensing optical depths.  This makes this technique ideal for tracing
the mass density profiles of elliptical galaxies over a broad redshift range
and for probing their evolution with cosmic time.

Samples of strongly lensed galaxies are further enriched by the, to some
extent complementary, imaging surveys (Cabanac et al. 2007; Faure et al.
2008; Kubo \& Dell' Antonio 2008; Ruff et al. 2011) which look for arc-like
features, and by radio surveys (Browne et al. 2003).  All that holds the
promise of a fast increase of the number of known strongly lensed sources,
fostered by the forthcoming large area optical (e.g., Oguri \& Marshall 2010)
and radio (SKA, Square Kilometer Array) surveys (e.g., Koopmans, Browne, \&
Jackson 2004). A simple, efficient, analytical tool applicable to the
analysis of large samples of galaxy-scale lenses is therefore warranted.

In this paper we work out exact and approximate solutions of the lens
equation based on a realistic model for the mass density profile of the lens
(\S\,\ref{sect:lens_eq}) and exploit them to reckon the lensing probability
as a function of the source redshift (\S\,\ref{sect:lens_prob}). As an
application of these results, following on the study of the high-$z$
luminosity function of galaxies measured by the H-ATLAS survey (Lapi et al.
2011), we compute, in \S\,\ref{sect:counts}, the number counts of  strongly
lensed sub-mm galaxies  implied by the physical model of galaxy formation and
evolution formulated by Granato et al. (2004) and further developed by Lapi
et al. (2006) and Mao et al. (2007). The model counts are compared with the
observational estimates by Negrello et al. (2010) and by Gonzalez-Nuevo et
al. (2012). While our study is focussed on early-type lenses, assumed to be
circularly symmetric, the cases of late-type lenses, of ellipticity, of the
presence of a super-massive black hole in the galactic nucleus, and of
super-galactic structures are discussed in \S\,\ref{sect:disc}. Finally, our
main results are summarized in \S\,\ref{sect:concl}.

Throughout the paper we adopt the standard flat $\Lambda$CDM cosmology (see
Komatsu et al. 2011) with current matter density parameter $\Omega_{\rm
M}=0.27$ and Hubble constant $H_0=72$ km s$^{-1}$ Mpc$^{-1}$.

\section{Lensing cross section}\label{sect:lens_eq}

\subsection{Lens mass models}

We focus here on galaxy-scale lensing, i.e., on those lensing events
where the deflector is a single/isolated early-type galaxy. The discussion of
the effect of the more extended dark matter (DM) halo of a group or cluster
in which the lens galaxy may reside, or of the disk of later type galaxies is
deferred to \S\,\ref{sect:disc}.

More specifically, we assume that the lens galaxy is associated to a DM halo
of mass $M_{\rm H}$ in the range $10^{11.4}-10^{13.5}\, M_\odot$
virialized at redshift $z_{\ell,v}\ga 1.5$. The redshift and the lower mass
limit are crudely meant to single out galactic halos associated with
individual spheroidal galaxies. Disk-dominated (and irregular) galaxies are
primarily associated with halos virializing at $z_{\ell,v}\la 1.5$, which may
have incorporated halos less massive than $10^{11.4}\, M_{\odot}$ virialized
at earlier times, that form their bulges. The upper mass limit to individual
galaxy halos comes from weak-lensing observations (e.g., Kochanek \& White
2001; Kleinheinrich et al. 2005), kinematic measurements (e.g., Kronawitter
et al. 2000; Gerhard et al. 2001), and from a theoretical analysis on the
velocity dispersion function of spheroidal galaxies (Cirasuolo et al. 2005).
The same limit is also suggested by modeling of the spheroids mass function
(Granato et al. 2004), of the quasar luminosity functions (Lapi et al. 2006),
and of the sub-mm galaxy number counts (Lapi et al. 2011).

For the DM we adopt a standard NFW (Navarro, Frenk, \& White 1996) profile
(e.g., Lokas \& Mamon 2001)
\begin{equation}
\rho_{\rm H}(r)={M_{\rm H}\over 4\pi R_{\rm H}^3}\, {f_c\, c^2\over (r/R_{\rm
H})\,(1+c\,r/R_{\rm H})^2},
\end{equation}
where $c$ is the concentration parameter and $f_c\equiv
[\log(1+c)-c/(1+c)]^{-1}$. The halo virial radius $R_{\rm H}$ is given by
(e.g., Bryan \& Norman 1998; Barkana \& Loeb 2001)
\begin{equation}
R_{\rm H} = 62\, \left({M_{\rm H}\over 10^{12}\, M_\odot}\right)^{1/3}\,
\left[{\Omega_{\rm M}\over \Omega_{\rm M}(z_{\ell,v})}\,
{\Delta_v(z_{\ell,v})\over 18\pi^2}\right]^{-1/3}\, \left({1+z_{\ell,v}\over
3.5}\right)^{-1}~\mathrm{kpc},
\end{equation}
in terms of the virialization redshift of the lens $z_{\ell,v}$, of the
evolved density parameter $\Omega_{\rm M}(z)=\Omega_{\rm M}\,
(1+z)^3/[\Omega_{\rm M}\, (1+z)^3+1-\Omega_{\rm M}]$, and of the overdensity
threshold for virialization $\Delta_v(z)=18\,\pi^2+82\,[\Omega_{\rm
M}(z)-1]-39\,[\Omega_{\rm M}(z)-1]^2$. For example, $z_{\ell,v}=2.5$ and
$M_{\rm H}=10^{13}\, M_{\odot}$ correspond to a virial size $R_{\rm H}\approx
200\,$kpc. Since the NFW profile yields a logarithmically diverging mass, we
set the edge of the halo at $R_{\rm H}$.

Hereafter we adopt $z_{\ell,v}=2.5$ as our fiducial value. This
choice is motivated on considering that massive early-type galaxies at
$z_\ell\sim 0.7$ feature relatively old ages $\sim 3-5$ Gyr of their stellar
populations, and formed the bulk of their stars over a timescale of order $1$
Gyr (for a review, see Renzini 2006; also Lapi et al. 2011). These evidences
point toward a virialization redshift of the host halo in the range
$z_{\ell,v}\sim 1.5-3.5$, consistent with the distribution of creation
redshifts found in numerical simulations for the massive halos considered
here (e.g., Moreno et al. 2009). Note that for an early-type lens the
observation redshift ($z_\ell\sim 0.7$; Gonz\'alez-Nuevo et al. 2012) is,
generally, substantially lower than the virialization one ($z_{\ell,v}\sim
2.5$). Therefore the frequently made approximation $z_\ell\approx z_{\ell,v}$
leads to a large overestimate of the halo size and, indirectly, to an
underestimate of the lensing probability.

Numerical simulations indicate that the concentration $c$ depends on halo
mass and redshift as (Prada et al. 2011)
\begin{equation}\label{eq:c}
c\approx 5\,\left({M_{\rm H}\over 10^{13}\, M_\odot}\right)^{-0.074}\,
\left({1+z_\ell\over 1.7}\right)^{-1},
\end{equation}
with a scatter of about $20\%$. For a lens with $M_{\rm H}=10^{13}\, M_\odot$
at redshift $z_\ell=0.7$ we have $c\approx 5$, that we adopt as our fiducial
value.

The DM to baryon ratio $M_{\rm H}/M_\star$ in early-type galaxies is
generally in the range $10-70$. In fact, this quantity can be roughly bounded
from below by the cosmological DM to baryon mass ratio (see Komatsu et al.
2011) that takes on values around $6$, and from above by the DM to stellar
mass ratio that statistical arguments (see Shankar et al. 2006; Lagattuta et
al. 2010; Moster et al. 2010) estimate to be around $70$. We take $M_{\rm
H}/M_\star=30$ as our fiducial value.

The impact on the lensing probability of different choices for
$z_{\ell,v}$, for $c$ [including the use of the halo-mass dependent
expression of Eq.~(\ref{eq:c})], and for $M_{\rm H}/M_\star$ is discussed in
\S\,\ref{sect:lens_prob}.

For the stellar component we adopt the 3-D S\'ersic profile (Prugniel \&
Simien 1997)
\begin{equation}
\rho_\star(r)={M_\star\over 4\pi R_e^3}\,{b_n^{2n}\over n\Gamma(2n)}\,
\left({r\over R_e}\right)^{-\alpha_n}\, e^{-b_n\,(r/R_e)^{1/n}},
\end{equation}
where $R_e$ is the effective radius, $n$ is the S\'ersic index, $b_n\equiv
2\,n-1/3+0.009876/n$, and $\alpha_n\equiv 1-1.188/2\,n+0.22/4\,n^2$.

The effective radius $R_e$ is related to the stellar mass by (Shen et al.
2003\footnote{Note that there is a typo in the normalization factor given in
Table~1 of Shen et al. (2003). The correct value is given in Cimatti et al.
(2008).}; Hyde \& Bernardi 2009)
\begin{equation}\label{eq:Re}
R_e\approx 1.28\, \left({M_\star\over 10^{10}\,M_\odot}\right)^{0.55}~\mathrm{kpc},
\end{equation}
found to hold (with a $\approx 30\%$ scatter) for $z_\ell\la 1$. Several
recent observational studies have found that massive, passively evolving
galaxies at $z\ga 1$ are much more compact than local galaxies of similar
stellar mass (Fan et al. 2010 and references therein). The study by Maier et
al. (2009), with high spectroscopic completeness, finds however that the size
evolution at fixed mass is modest ($\sim 25\%$) from $z\sim 0.7$ to $z=0$,
i.e. up to our reference value of the lens redshift. We have checked
that decreasing $R_e$ at fixed $M_\star$ by a factor $f_e>1$ increases
somewhat the probability of amplifications only in the range $3\la \mu \la
5$; the effect becomes essentially independent of the decrease factor for
$f_e \ga 1.5$.

Values of the S\'ersic index for massive early-type galaxies are in the range
$n\approx 3-10$, with a tendency for more massive systems to feature higher
values (e.g., Kormendy et al. 2009). Early-type galaxies with $n=2$ generally
are either dwarf spheroidals or contain a substantial disk component and do
not obey Eq.~(\ref{eq:Re}).  We will consider a fiducial value $n=4$,
corresponding to the classical de Vaucouleurs (1948) profile. Again, the
effect on the lensing probability of different values is investigated in
\S\,\ref{sect:lens_prob}.

To sum up, we adopt a two-component model, made of a stellar component with a
S\'ersic profile plus a DM halo with a NFW profile. Given the halo mass
$M_{\rm H}$ and the virialization redshift, $z_{\ell,v}$, the total mass
distribution of the lens galaxy is specified by the parameters $z_\ell$,
$M_{\rm H}/M_\star$, and $n$ [after Eq.~(\ref{eq:c}) $c$ is determined by
$M_{\rm H}$ and $z_\ell$]. Hereafter this two-component model will be
referred to as the `SISSA model'\footnote{From the name of our main
institution. The acronym `SISSA' and `SIS' sound as close as the
corresponding model results are.}. The lensing probability distribution
yielded by this model will be compared with those yielded by two other
commonly used models. The first (hereafter referred to as `NFW model') just
consists in adopting a pure NFW profile, hence neglecting the effect of the
baryons. The second (hereafter referred to as the `SIS model') adopts the
classic singular isothermal sphere density profile
\begin{equation}
\rho_{\rm SIS}(r)={\sigma_{\rm SIS}^2\over 2\pi\,G}\,{1\over
r^2} = {M_{\rm H}\over 4\pi R_{\rm H}^3}\, \left({r\over R_{\rm
H}}\right)^{-2},
\end{equation}
where $\sigma_{\rm SIS}$ is the 1-D velocity dispersion of the overall mass.
The second equality follows from the commonly used assumption $\sigma_{\rm
SIS}\approx V_{\rm H}/\sqrt{2}$, in terms of the halo circular velocity
$V_{\rm H}^2=G\, M_{\rm H}/R_{\rm H}$.

\subsection{Surface density profile}

The surface density writes
\begin{equation}
\Sigma(s)=\int{\mathrm d}r~{r\over \sqrt{r^2-s^2}}\, \rho(r),
\end{equation}
$s$ being the radial coordinate projected on the plane of the sky. It is
important to remember that the surface density becomes effective for
\emph{strong} lensing when it exceeds the critical threshold
\begin{equation}\label{eq:Sigma_c}
\Sigma_c={c^2\over 4\pi\,G}\,{(1+z_\ell)\, D_s\over D_\ell\, D_{\ell s}}
\end{equation}
corresponding to a convergence $\kappa=1$ for a thin lens. Here $D_s$,
$D_\ell$, and $D_{\ell s}=D_s-D_\ell$ are the \emph{comoving} angular
distances (also called proper motion distances; see Kochanek 2006) from the
source at $z_s$ to the observer at $z=0$, from the lens at $z_\ell$ to the
observer at $z=0$, and from the source at $z_s$ to the lens at $z_\ell$,
respectively. In a flat Universe the comoving angular distances are defined
as $D_{ij}\equiv(c/H_0)\, \int_{z_i}^{z_j}{\rm
d}z\,[\Omega_M\,(1+z)^3+1-\Omega_M]^{-1/2}$. The angular diameter distance is
$D_{ij}/(1+z_j)$ and the luminosity distance is $D_{ij}\,(1+z_j)$.

\begin{figure}
\epsscale{0.8} \plotone{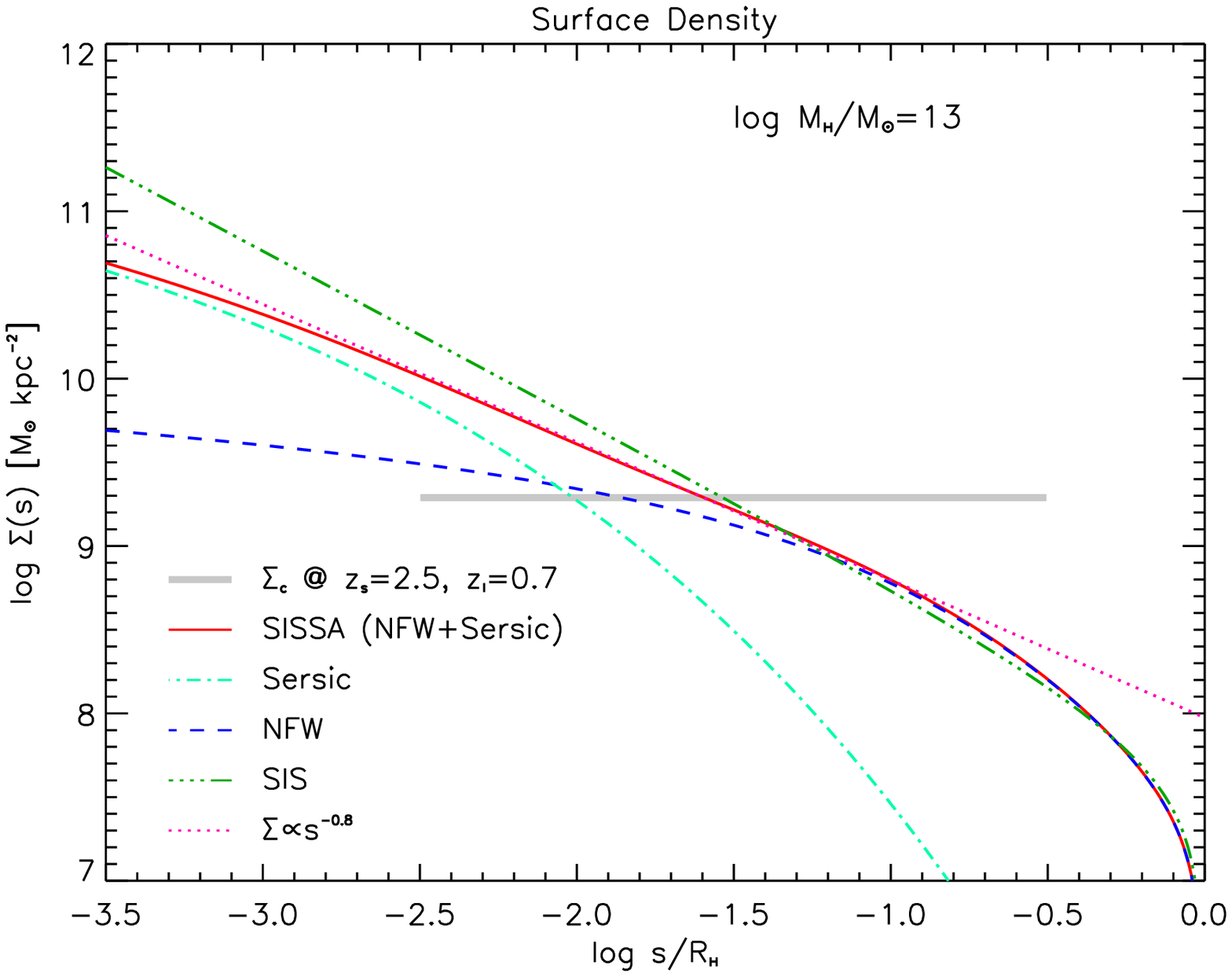} \caption{Surface mass density of an
early-type lens. (Blue) dashed line: NFW dark matter profile with $M_{\rm
H}=10^{13}\, M_{\odot}$ and concentration parameter $c=5$; (cyan) dot-dashed
line: S\'ersic profile ($n=4$) of the stellar component in the proportion
$M_{\rm H}/M_\star=30$; (red) solid line: SISSA model constituted by the sum
of the two contributions; (green) triple-dot-dashed line: classical SIS model
for the same DM mass; (magenta) dotted line: power law relation
$\Sigma(s)\propto s^{-0.8}$, that provides a good approximation to the SISSA
model in the radial range relevant for strong gravitational lensing (see text for
details). The horizontal grey line represents the critical density for
lensing for a source redshift $z_s=2.5$ and a lens redshift $z_\ell=0.7$. The
projected radius $s$ is normalized to the halo virial
radius.}\label{fig:surface_dens}
\end{figure}

In Fig.~\ref{fig:surface_dens} the surface mass density yielded by the SISSA
model for a lens at $z_\ell=0.7$ with $M_{\rm H}=10^{13}\, M_\odot$,
$z_{\ell,v}=2.5$, $c=5$, $n=4$, and $M_{\rm H}/M_\star=30$, is compared with
those yielded by the S\'ersic, NFW and SIS laws. The horizontal grey line
represents the critical surface density when the source is at $z_s=2.5$ and
the lens at $z_\ell=0.7$. It shows that, in this case, only the matter
located at $s\la 10^{-1.5}\, R_{\rm H}$ is effective for strong lensing. The
stellar contribution to the surface density dominates for $s \la 10^{-2}\,
R_{\rm H}$, while the DM contribution takes over at larger radii.

As illustrated by Fig.~\ref{fig:surface_dens} in the radial range $-2.5\la
\log(s/R_{\rm H})\la -1$, which generally contributes most to the
gravitational deflection, the combination of the stellar and DM components to
the total surface density closely mimics a power law
\begin{equation}\label{eq:power_law}
\Sigma(s)=\Sigma_0\,\left({s\over s_0}\right)^{-\eta}.
\end{equation}
In Table~\ref{tab:fits} we show that, at fixed halo mass, $M_{\rm H}$, both
the normalization $\Sigma_0$ at the reference radius $s_0\approx 10^{-2}\,
R_{\rm H}$, and the power-law index $\eta\approx 0.8-0.9$ are only weakly
dependent on the parameters of the mass distribution. Specifically, $\eta$
and $\Sigma_0$ slightly increase with decreasing $M_{\rm H}/M_\star$ (i.e.
for higher stellar contributions in the inner region), with increasing halo
concentration (i.e. for a higher DM contribution in the inner region), and
with decreasing S\'ersic index (i.e. for a stellar distribution more
concentrated at the center). At fixed parameters of the lens mass
distribution, $\Sigma_0$ (obviously) decreases while $\eta$ increases with
decreasing halo mass.

The slopes of the power-law approximation are in good agreement with those
determined from the stellar dynamics (e.g., Thomas et al. 2011), from the
globular cluster/planetary nebulae kinematics (e.g., Rodionov \& Athanassoula
2011), from the HI gas disk/ring (e.g., Weijmans et al. 2008), from the
profile of the X-ray emission (e.g., Churazov et al. 2010; Humphrey \& Buote
2010), and from gravitational lensing in individual galaxies (e.g., Koopmans
et al. 2009; Spiniello et al. 2011; Barnab\'e et al. 2011). All these
observational determinations concur in indicating that the overall density
profile is roughly isothermal (at least in the inner region), with surface
density slopes around $\eta\approx 1$.

In comparing the observational determinations of the slopes of the density
profiles it must be kept in mind that the profiles are not real power laws.
This implies that the slope of the volume density, $\eta_V$, is not simply
related to the slope of the surface density, $\eta$, by $\eta_V=\eta+1$, as
in the power-law case. For the combination of S\'ersic and NFW profiles
considered here we find that the best fit slopes over the radial range
relevant for strong lensing are related by $\eta_V\approx \eta+1.2$, due to
projection effects. Thus, the slightly under-isothermal values of $\eta$
yielded by the SISSA model (see Table~\ref{tab:fits}) are fully consistent
with the slightly over-isothermal values of $\eta_V$ (in the range $1.9-2.3$)
found by Koopmans et al. (2009) and Barnab\'e et al. (2011). The latter
authors also find hints of a steeper slope $\eta$ for the least massive
systems, consistent with the trend coming out of Table~\ref{tab:fits}.
Finally, note that during the galaxy lifetime the mass density
profile may evolve under the influence of various physical processes; we
address the issue in \S~\ref{sect:evolution}.

\subsection{Lens equation: exact and approximated solutions}\label{sect:solutions}

In the following we consider circular lenses, deferring to
\S\,\ref{sect:disc} the discussion on the effect of  ellipticity. In such a
case, the light rays coming from a distant, point-like source are deflected
by an angle
\begin{equation}\label{eq:alpha}
\alpha(\theta|z_\ell,z_s)={2\over \theta}\, \int_0^\theta {\rm d}\vartheta\,
\vartheta\, {\Sigma(s)\over \Sigma_c}={4 G\, M(<\xi)\over c^2\, \xi}\,
{D_{\ell s}\over D_s},
\end{equation}
where $\theta$ is the angular distance of the image from the lens center,
$\vartheta\equiv s\,(1+z_\ell)/D_\ell$, and $M(<\xi)\equiv
2\pi\int_0^\xi {\rm d}s\, s\, \Sigma(s)$ is the mass within the
projected radius $\xi=\theta D_\ell/(1+z_\ell)$.

The relation between the position of the source and of its (possibly
multiple) lensed image(s) relative to the observer is described by the lens
equation
\begin{equation}
\beta=\theta-{\theta\over |\theta|}\, \alpha(|\theta|).
\end{equation}
Here $\beta$ and $\theta$ are the angles formed by the source and by its
images with the optical axis, i.e. with the imaginary line connecting the
observer and the center of the lens mass distribution. Solving the lens
equation means finding all the positions $\theta$ of the images for a given
source position $\beta$.

The amplification of the images can be computed as:
\begin{equation}
\mu={1\over \lambda_+\, \lambda_-}~~~~\mathrm{with}~~\lambda_+\equiv 1-{{\rm
d}\alpha\over {\rm d}\theta}~~\mathrm{and}~~\lambda_-\equiv 1-{\alpha\over
\theta}.
\end{equation}
If either of the two quantities $\lambda_\pm$ vanishes, the magnification
formally diverges. Thus the condition $\lambda_\pm=0$ defines {\it critical}
curves in the lens plane and corresponding {\it caustics} in the source
plane. The magnification can be positive or negative, implying that the image
has positive or negative parity or, equivalently, is direct or reversed. The
total magnification $\mu_{\rm tot}$ is the sum of the absolute values of the
magnifications for all the images, i.e., ignoring parity.

A point-like source perfectly aligned with the observer and the center of the
foreground mass distribution is lensed into a ring of radius $\theta_E$,
called Einstein ring. For a circular lens it coincides with the
critical curve defined by $\lambda_-=0$. The other critical
curve defined by $\lambda_+=0$, if present, approximately coincides with an
inner ring of radius $\theta_I < \theta_E$. In Fig.~\ref{fig:solutions} we
show an example of the numerical solutions of the lens equation for the SISSA
model. The source is located at redshift $z_s=2.5$ and the lens at redshift
$z_\ell=0.7$. The parameters of the lens mass distribution are as in
Fig.~\ref{fig:surface_dens}. The Einstein ring is located at $\theta_E\approx
1.82^{\prime\prime}$. In addition, the SISSA model (and also the NFW model,
but not the SIS model) features another, inner critical curve at
$\theta_I\approx 0.32^{\prime\prime}$. In the top panel we illustrate the
relation between the angular positions of the source $\beta$ and of the
images $\theta$, normalized to the position $\theta_E$ of the Einstein ring.
In the bottom panel, we illustrate the relation between the angular position
of the source $\beta$ and the amplification of the different images,
including the total magnification summed over them.

As it can be seen from the bottom panel of Fig.~2, the SISSA model
features one image for $\beta\ga \beta_I$ and three images for $\beta\la
\beta_I$, where $\beta_I\approx 0.5\, \theta_E$ is the location of the
outer caustic corresponding to the inner critical curve at
$\theta_I$. Of the three images, however, the one closest to the optical
axis (dot-dashed line) is strongly demagnified, while the other two
(solid and dashed lines) are amplified by different amounts and
feature opposite parity. At the location of the outer caustic
$\beta=\beta_I$, the second and third images are degenerate with infinite
magnification (in the plot the red solid curve referring to the total
magnification should go to $\mu\rightarrow \infty$ at $\beta=\beta_I$, but
the divergence has been smoothed out for clarity).

In Fig.~\ref{fig:angles} we illustrate how the angular positions of the
Einstein ring $\theta_E$ and of the inner critical curve $\theta_I$
depend on the lens mass and redshift. In the top panel we focus on the SISSA
model, and show how the position of the Einstein ring and of the inner
critical curve vary as a function of the lens redshift $z_\ell$, for
different source redshifts $z_s$. The bottom panel elucidates the dependence
of $\theta_E$ and $\theta_I$ on the lens mass for different lens models, at
fixed $z_s=2.5$ and $z_\ell=0.7$. Plainly, $\theta_E$ and $\theta_I$ increase
with the halo mass since, after Eq.~(\ref{eq:alpha}), more mass implies wider
bend angles. The SISSA model yields larger values of $\theta_E$ than the NFW
and SIS models but smaller values of $\theta_I$ than the NFW model (the SIS
model has no inner critical curve).

These behaviors, and the comparisons among different mass models, may be
better understood looking at approximate solutions of the lens equations
obtained using the power-law description of the projected mass density
[Eq.~(\ref{eq:power_law})] that, as shown by Fig.~\ref{fig:surface_dens},
approximates quite well a realistic density profile in the radial range most
relevant for strong lensing. For the SISSA model, $\eta\approx 0.8-0.9$ is
weakly dependent on the parameters of the mass distribution (see
Table~\ref{tab:fits}). The SIS model has a slightly steeper slope ($\eta_{\rm
SIS}= 1$), while, in the range most relevant for strong lensing, the NFW can
be approximated with a flatter slope $\eta_{\rm NFW}\approx 0.3-0.4$.

Under the power law approximation the deflection angle due to the lens
potential within a circle of angular radius $\theta$ is
\begin{equation}
\bar \alpha(\theta)=|\bar \theta|^{1-\eta}.
\end{equation}
Here and in the following the over-bar (e.g., $\bar \theta\equiv
\theta/\theta_E$) denotes normalization to the Einstein radius, that for
power-law lens models can be simply written as
\begin{equation}\label{eq:thetaE}
\theta_E=\theta_0\, \left({2\over 2-\eta}\, {\Sigma_0\over
\Sigma_c}\right)^{1/\eta},
\end{equation}
with $\theta_0\equiv s_0\,(1+z_\ell)/D_\ell$ [cf. Eq.~(9)].
Table~\ref{tab:fits} shows that, at fixed lens and source redshift, the
factor $[2\,\Sigma_0/(2-\eta)\,\Sigma_c]^{1/\eta}$ scales approximately as
$M_{\rm H}^{1/3}$, implying $\theta_E\propto M_{\rm H}^{2/3}$ since trivially
$\theta_0\propto R_{\rm H}\propto M_{\rm H}^{1/3}$. This illustrates the
power of lensing for weighing the halos. Since
\begin{equation}
\lambda_-=1-|\bar\theta|^{-\eta}~~~~~~~~\lambda_+=1-(1-\eta)\,|\bar\theta|^{-\eta}
\end{equation}
the magnification of an image is
\begin{equation}\label{eq:mu}
\mu\equiv {1\over \lambda_+\,\lambda_-}={1\over
[1-|\bar\theta|^{-\eta}]\,[1-(1-\eta)\,|\bar\theta|^{-\eta}]}.
\end{equation}
This equation highlights that, in addition to the critical curve
corresponding to the Einstein ring ($\theta=\theta_E$), for $\eta<1$ there is
also an inner ring at
\begin{equation}
\theta_I=\theta_E\, (1-\eta)^{1/\eta}.
\end{equation}
To get the magnification from Eq.~(\ref{eq:mu}) we need to compute the
positions, $\bar\theta$, of the images as a function of the angular distance
$\bar \beta$ (in units of $\theta_E$) of the source from the optical axis by
solving the lens equation
\begin{equation}
\bar \beta=\bar \theta\, -{\bar\theta\over
|\bar\theta|}\,|\bar\theta|^{1-\eta}.
\end{equation}
Summing up over all the images yields the global magnification of the lens
model as a function of $\beta$.

For the SIS model ($\eta=1$) this gives a constant deflection
$\alpha=\theta_E$, and no radial critical curve. For $\bar \beta>1$,
i.e., outside $\theta_E$, the lens equation yields only one image at $\bar
\theta=1+\bar\beta$ with magnification $\mu=1+1/\bar\beta$. On the other
hand, for $\bar \beta\la 1$, i.e., inside $\theta_E$, it gives the two
images $\bar \theta_\pm=\bar \beta\pm 1$, and related magnifications
$\mu_\pm=1\pm 1/\bar\beta$; thus the total magnification amounts to
$\mu=|\mu_+|+|\mu_-|=\mu_+-\mu_-=2/\bar\beta$.

For a generic $\eta<1$ it is not possible to solve the lens equation
analytically but we find that the numerical solutions can be well
approximated, over the amplification range $1\la \mu\la 30$ and over the
range of parameters explored in Table~\ref{tab:fits}, by the expressions
\parbox{10cm}{
\begin{eqnarray}
\nonumber\bar\beta=&1/(\mu-1)^\eta~~~~~~&\mathrm{for}~~~\bar\beta\geq \bar\beta_I\\
\nonumber\bar\beta=&(2/\mu)^\eta~~~~~~&\mathrm{for}~~~\bar\beta\leq \bar\beta_I,
\end{eqnarray} } \hfill
\parbox{1cm}{\begin{eqnarray}\end{eqnarray}}

\noindent that recover the SIS solutions for $\eta=1$. The value
$\bar\beta_I\equiv \eta\,(1-\eta)^{-1+1/\eta}$ corresponds to the location of
the inner critical curve $\theta_I$.

The goodness of these approximations may be appreciated by eye on looking at
the dotted lines in Fig.~\ref{fig:solutions} (both panels). They can be
useful not only for fast computations of the lensing cross section, but also
for constructing simulated lensed images of extended sources via the
ray-tracing technique or, inversely, for reconstructing the lens mass profile
from images of lensed sources.

\subsection{Cross sections}\label{sect:cross_sec}

Given the relation between the source position $\beta$ and the total
magnification of the images $\mu$, the cumulative cross section for lensing
magnification, as a function of the lens halo mass $M_{\rm H}$ and of the
lens and source redshifts $z_\ell$ and $z_s$, simply writes:
\begin{equation}
\sigma(>\mu,M_{\rm H},z_s,z_\ell)=\pi\, \beta^2(\mu).
\end{equation}
In Fig.~\ref{fig:cross_sect} we illustrate the cumulative lens cross section
for the SISSA, NFW and SIS models as a function of $M_{\rm H}$, for
$z_s=2.5$, $z_\ell=0.7$ and two values of $\mu$. For massive halos moderate
amplifications ($\mu\la 5$) are mainly contributed by the outer regions of
the mass distribution where the DM dominates. Thus the SISSA and the NFW
models yield very similar values of the cumulative cross section for $2< \mu
<5$. On the other hand, strong amplifications ($\mu>10$) are mainly
contributed by inner regions where the stellar component becomes important or
even dominant. Correspondingly, the SISSA cross section is considerably
higher than the NFW and close to the SIS one. The different shapes of the
cross sections as a function of halo mass mainly reflect the different
dependencies of $\theta_E$ on $M_{\rm H}$ for the three models. The value of
the projected radius $s_c$ at which the surface density equals the critical
threshold for lensing, $\Sigma_c$ [Eq.~(\ref{eq:Sigma_c})], decreases with
decreasing $M_{\rm H}$, causing a decrease of the lensing cross section. In
the case of the NFW model, below $\log(M_{\rm H}/M_\odot)=12$ the radius
$s_c$ lies in the region where the profile is much flatter than that of the
SIS and SISSA models. As a consequence, it decreases faster with decreasing
$M_{\rm H}$ and, correspondingly, the lens cross section drops.

\subsection{Lensing by extended sources}

The above results apply to the idealized case of a point-like source. What
about extended sources? The problem can be solved via the ray-tracing
technique (e.g., Schneider, Ehlers \& Falco 1992, p. 304), i.e. by applying
the $\beta-\theta$ and $\beta-\mu$ relations to every point of the unlensed
light distribution of the source. Some examples are shown in
Figs.~\ref{fig:ray_tr1} to \ref{fig:ray_tr3} where $z_s=2.5$ and
$z_\ell=0.7$. In the cases of Figs.~\ref{fig:ray_tr1} and \ref{fig:ray_tr2}
the light distribution of the source is modeled as a standard S\'ersic
profile with $n=4$, while Fig.~\ref{fig:ray_tr3} pertains to a galaxy
comprising multiple (4 in this example) luminous, giant "clumps" that appear
to be a ubiquitous feature of high-redshift star-forming galaxies (e.g.,
F{\"o}rster Schreiber et al. 2011). The results are shown for different mass
models.

Figure~\ref{fig:ray_tr1} refers to a zero impact parameter, i.e. the center
of the source is aligned with the optical axis. The central regions of the
source are strongly amplified and deformed into the Einstein ring. The NFW
and SISSA models (but not the SIS model) yield also an inner critical
curve, difficult to see because of the limited resolution of the figure. In
Fig.~\ref{fig:ray_tr2} the impact parameter is small but non-null. The
Einstein ring and the inner critical curve are still present since
there are regions of the unlensed extended source that are crossed by the
optical axis. But the central, brightest region is lensed into two deformed
and amplified images, one inside and the other outside the Einstein ring. In
Fig.~\ref{fig:ray_tr3} the 4 clumps have small impact parameters (specified
in the caption). The lensed image looks like an Einstein ring with a knotty
structure. High resolution imaging would be necessary to distinguish this
configuration from that produced by a smooth source profile with an impact
parameter close to zero.

This analysis also provides an useful way to estimate the maximum
amplification $\mu_{\rm max}$ to be expected from extended sources, when
their multiple images in the lens plane are not resolved (e.g., Peacock
1982). We compute the quantity $\mu_{\rm max}$ as the ratio between the total
flux of the lensed images to that of the unlensed source over the image plane (see
Perrotta et al. 2002 for details). In Fig.~\ref{fig:max_ampl} we plot
$\mu_{\rm max}$ versus the offset between the center of the extended source and
the optical axis for the SISSA (red line), NFW (blue), and SIS (green) lens
models. The outcomes are quite similar, with the SISSA model providing
slightly higher maximum amplifications. For the  spherical lenses considered
here $\mu_{\rm max}$ increases monotonically with decreasing offset.

The figure also illustrates, for the SISSA model, the dependence of $\mu_{\rm
max}$ on the half stellar mass radius, $R_e$, of the source. The dissipative
collapse of baryons within the DM halos can result in $R_e \approx 1-3$ kpc
for $\log(M_{\rm H}/M_\odot)= 12-13$ (Fan et al. 2010). The corresponding
values of $\mu_{\rm max}$ for close alignments between the source and the
lens are in the range $30-50$; $\mu_{\rm max}$ depends inversely on the
source size, and decreases to $\approx 10$ for $R_e=10\,$kpc. Thus the
magnification distribution of sub-mm galaxies can provide information on the
scale of the stellar mass distribution for dusty high-$z$ galaxies, difficult
to determine by other means. We also find that $\mu_{\rm max}$ increases
linearly with the source angular diameter distance, which is however very
weakly dependent on redshift in the range of interest here ($1.5 \la z_s \la
4$).

\section{Lensing probabilities and distributions}\label{sect:lens_prob}

We compute the lensing optical depth as
\begin{equation}\label{eq:tau}
\tau(z_s|>\mu)=\int_0^{z_s}{\rm d}z_\ell~\int{\rm d}M_{\rm H}~{{\rm
d}^2N\over {\rm d}M_{\rm H} {\rm d}V}\, {{\rm d}^2 V\over {\rm d}z_\ell {\rm
d}\Omega}~\sigma(>\mu,M_{\rm H},z_s,z_\ell),
\end{equation}
where ${\rm d}^2V/{\rm d}z {\rm d}\Omega$ is the comoving volume per unit
$z-$interval and solid angle, while ${\rm d}^2 N/{\rm d}M_{\rm H}\,{\rm d}V$
is the galaxy halo mass function (see Shankar et al. 2006 for details), i.e.
the statistics of halos containing one single galaxy. As in Lapi et al.
(2006) the galaxy halo mass function is computed from the standard Sheth \&
Tormen (1999, 2002) halo mass function by (i) accounting for the possibility
that a DM halo contains multiple subhalos each hosting a galaxy and (ii)
removing halos corresponding to galaxy systems rather than to individual
galaxies. We deal with (ii) by simply cutting off the halo mass function at a
mass of $10^{13.5}\,M_\odot$ beyond which the probability of having multiple
galaxies within a halo quickly becomes very high (e.g., Magliocchetti \&
Porciani 2003). As for (i), we add the subhalo mass function, following the
procedure described by Vale \& Ostriker (2004, 2006) and Shankar et al.
(2006), and using the fit to the subhalo mass function at various redshifts
provided by van den Bosch, Tormen \& Giocoli (2005). However, we have checked
that for the masses and redshifts relevant here ($z_{\rm \ell,v}\ga 1.5$ and
$11.4\la \log(M_{\rm H}/M_\odot)\la 13.5$), the total (halo + subhalo) mass
function differs from the halo mass function by less than $5\%$. Note that in
the computation of Eq.~(\ref{eq:tau}) we do not include the contribution to
lensing by massive groups and clusters, for which the parameters of the lens
mass distribution are different from those adopted here (see discussion in
\S\,\ref{sect:disc}).

As illustrated by Fig.~\ref{fig:tau}, the lensing optical depth increases
very rapidly with increasing source redshift up to $z_s\approx 1$, grows by a
factor of $\approx 4$ between $z_s=1$ and $z_s=2$ and by a further factor
$\approx 2.5$ at $z_s=5$. When the magnification threshold increases from
$\mu=2$ to $\mu=10$, $\tau(z_s|>\mu)$ decreases by factors of $\approx
25-30$ for the SIS and SISSA models and by a much larger factor ($>100$) for
the NFW model.

The inner integral in Eq.~(\ref{eq:tau}) gives the lens redshift distribution
${\rm d}p(z_\ell|>\mu,z_s)/{\rm d}z_\ell$, i.e. the surface density per unit
redshift interval of lenses located at $z_\ell$ that can produce a strong
lensing event with total magnification $>\mu$ on a source at redshift $z_s$.
Examples of lens redshift distributions for the NFW, SIS and SISSA models,
with an amplification threshold $\mu=2$ are shown in Fig.~\ref{fig:zdistr}.
They are similar with broad peaks at $z_\ell\approx 0.6-0.7$ for $z_s=2.5$.
However the SIS and SISSA models yield higher high-$z$ tails than the NFW
model. As illustrated for the SISSA model, as the source redshift, $z_s$,
increases so does the peak of the $z_\ell$ distribution, the peak broadens
and the high-$z$ tail becomes more prominent.

The differentiation of Eq.~(\ref{eq:tau}) with respect to $\mu$ yields
(minus) the differential magnification distribution ${\rm d}p(\mu|z_s)/{\rm
d}\mu$, illustrated in Fig.~\ref{fig:amplif_distr} for the NFW, SIS, and
SISSA models with $z_s=2.5$. The distributions are similar below $\mu\approx
3$. For higher magnifications the SIS and SISSA models are close to each
other and increasingly above the NFW one. The only appreciable difference
between SIS and SISSA occurs in the range $3<\mu<6$ corresponding to the
transition between magnifications dominated by the DM to magnifications
dominated by the stellar component. As visualized in
Fig.~\ref{fig:amplif_distr} for the SISSA model, increasing $z_s$ increases
the normalization of the amplification distribution, reflecting the increase
in the lensing optical depth, without substantially affecting its shape.

Figure~\ref{fig:parameters} elucidates the dependence of the SISSA
amplification distribution on the parameters of the lens mass model. Lower
$M_{\rm H}/M_\star$ ratios, i.e. larger amounts of stellar mass, yield higher
probability of large amplifications and lower values of the transition $\mu$
between the DM and the star dominated regime (the kink shifts to the left;
upper left-hand panel). Interestingly, taking into account the mass
dependence of the $M_{\rm H}/M_\star$ ratio implied by the Lapi et al. (2011)
galaxy formation model yields results similar to those obtained a constant
value $M_{\rm H}/M_\star=30$. The effect of varying the S\'ersic index of the
stellar component (upper right-hand panel) is small for $n\ge 4$: as $n$
increases we get slightly smaller probabilities of large amplifications and
slightly higher probabilities of low amplifications. Note that the trend with
decreasing $n$, and especially the feature around $\mu\approx 5$, cannot be
extrapolated to lower values of $n$ because Eq.~(\ref{eq:Re}) is no longer
appropriate to compute the effective radius $R_e$. In particular $n=1$
corresponds to exponential disks and disk galaxies have much larger effective
radii than given by Eq.~(\ref{eq:Re}). Correspondingly, they have much
smaller surface densities, hence much lower probabilities of large
amplifications. The lensing by spiral galaxies will be discussed in
\S\,\ref{sect:disc}.

The bottom left-hand panel shows that higher values of the concentration
parameter $c$ of the host DM halo yield higher probabilities of large
amplifications (coming from the gravitational field in the inner regions of
the lens) compensated by a tiny decrease (imperceptible in the figure) of the
probability of small amplifications. Using the mass dependent parametrization
of Eq.~(\ref{eq:c}) gives results almost indistinguishable from those
obtained using our fiducial value $c=5$. Finally, the bottom right-hand panel
shows that the probability of large amplifications increases with increasing
virialization redshift, $z_{\ell,v}$, of the lens, as expected since, for
given mass, both the halo radius and the stellar effective radius decrease;
correspondingly, the surface density increases. As a consequence, adopting
$z_{\ell,v}=z_\ell$, as done in some analyses, underestimates the probability
of large amplifications.

The distributions of angular separations of the two brightest images for our
fiducial lens properties (top panel of Fig.~\ref{fig:distributions}) are very
similar for the SIS and SISSA models. In both cases the distribution peaks at
$1.58''$ and has a FWHM of $3.58''$. The NFW model peaks at the same angular
separation but has a substantially higher probability of smaller angular
separations.  The dependence on the source redshift is very weak. Note that
we are considering only galaxy-scale lensing, i.e. we are not taking into
account lenses on super-galactic scales. The distribution of the differences
in the propagation times from the source to the observer for the two
brightest images, referred to as time delays, yielded by the SISSA model is
shifted towards slightly shorter values compared to the SIS model; the NFW
model implies still shorter delays, with a significant probability at $\Delta
t\la 1$ day (for details on the computation of time delays see, e.g.,
Porciani \& Madau 2000; Oguri et al. 2002). The dependence on the source
redshift is appreciable, with the peak of the distribution increasing as a
function of $z_s$ (see also Oguri et al. 2002; Li \& Ostriker 2003). This is
not in contradiction with the fact that the time delay at given $z_\ell$
decreases with $z_s$ as $D_s/D_{\ell s}$, since in computing the time delay
distribution the quantity $1/D_{\ell s}$ must be integrated over $z_\ell$ as
in Eq.~(21), and both the outcome of this integral and $D_s$ increase with
$z_s$.

\section{Number counts of lensed sub-mm galaxies}\label{sect:counts}

Given the number density ${\rm d}^2N/{\rm d}S {\rm d}z_s$ of unlensed sub-mm
galaxies per unit flux density and redshift interval and the amplification
distribution ${\rm d}p/{\rm d}\mu$ the observed counts allowing for the
effect of lensing can be computed as
\begin{equation}
{{\rm d}N\over {\rm d}\log S}(S)=\int{\rm d}z_s~\int{\rm d}\mu~{{\rm d}p\over
{\rm d}\mu}(\mu|z_s)\, {{\rm d}^2 N\over{\rm d}\log S {\rm d}z_s} (S/\mu).
\end{equation}
Here we have approximated to unity the factor $1/<\mu>$ that would have
appeared on the right-hand-side, as appropriate for large-area surveys (see
Jain \& Lima 2011). For the unlensed counts we adopt the model by Lapi et al.
(2011) that successfully reproduces the (sub-)mm counts from $250$ to
$1100\,\mu$m. The resulting Euclidean normalized counts of point-like sources
at $500\, \mu$m and $350\, \mu$m are shown in Fig.~\ref{fig:counts}; standard
parameters of the lens mass distribution have been adopted. The source
extension translates into an upper limit on the amplification
(Fig.~\ref{fig:max_ampl}) whose effect on the counts is set out in the lower
panel of Fig.~\ref{fig:zoom_counts350}. For compact sources with effective
radii in the range $R_e=1-3$ kpc, expected to be typical for the high-$z$
galaxies of interest here, the predicted counts are quite similar to those
for point-like sources. The model counts compare quite well with the counts
of confirmed strongly lensed galaxies selected at $500\,\mu$m by Negrello et
al. (2010) and with the counts of candidate strongly lensed galaxies selected
via the HALOS method at $350\,\mu$m by Gonzalez-Nuevo et al. (2012). Note
that we have corrected the SDP/HALOS counts with the flux-dependent purity
estimated by the latter authors (see their \S~4.3 and Fig.~10), which amounts
to about $70\%$ for flux densities $S_{350\,\mu{\rm m}} \ga 85$ mJy.

The completeness of this sample as a function of flux is difficult to
assess accurately. The main source of incompleteness is constituted by the
fact that faint (mostly high-redshift) lenses may have been missed by the
optical surveys exploited by Gonzalez-Nuevo et al. (2012). However, such
surveys are expected to be reasonably complete for the massive galaxies
($M_\star\ga 3\times 10^{11}\, M_\odot$) acting as lenses in the redshift
range $z_\ell\sim 0.5-0.8$ where the lensing probability peaks.

The upper panel of Fig.~\ref{fig:zoom_counts350} contrasts SISSA with SIS and
NFW models. While even with much larger samples it will be hard to
discriminate between the SISSA and SIS predictions, the NFW model gives
substantially lower counts at the brightest flux densities. Thus the counts
of bright strongly lensed sources over an H-ATLAS area ten times larger than
the SDP field (M. Negrello et al., in preparation) can provide a significant
test. The lower panel of Fig.~\ref{fig:zoom_counts350} illustrates the
contributions to the counts of different amplification intervals. Most of the
contribution at flux densities $S_{350\,\mu{\rm m}} < 80\,$mJy comes from
moderate amplifications ($2\la \mu\la 5$). Larger amplifications, up to
$\mu\approx 30$, become increasingly important at brighter flux densities.
Amplifications $\mu\ga 30$ are very rare and have little effect on the
counts, although a few extreme cases may show up in very large area surveys,
such as those made by the \textsl{Planck} satellite.

An important application of strong lensing is the possibility of
investigating sources fainter than those accessible with other means. In
particular, sub-millimeter surveys have proven to be an extremely powerful
tool to investigate the  dust enshrouded most active star formation phases of
galaxy evolution. However, the detected high-$z$ sub-mm sources are generally
intrinsically ultra-luminous galaxies, forming stars at extreme rates, (above
several hundreds solar masses per year; Lapi et al. 2011), and are therefore
atypical. Data (Genzel et al. 2006; F\"oster-Schreiber et al. 2006) suggest
that the most effective star formers in the universe are galaxies with
stellar and gas masses $\sim 10^{11}\,M_\odot$ at $z\approx 2-3$, endowed
with star-formation rates of SFR $\sim 100-200\, M_\odot$ yr$^{-1}$, that
generally have sub-mm flux densities below the confusion limits of large
(sub-)mm surveys (from ground-based single dish telescopes or from
\textsl{Herschel}). It is therefore interesting to investigate the flux
density range best suited to select such galaxies.
Figure~\ref{fig:counts350_sfr} shows that, thanks to strong lensing, the
H-ATLAS survey will allow us to sample galaxies down to SFR $< 200\,M_\odot$
yr$^{-1}$ already at relatively bright flux density limits, where the
selection of strongly lensed galaxies is easier (Negrello et al. 2010;
Gonz\'alez-Nuevo et al. 2012). For example, we expect $\approx 0.2$ strongly
lensed sources per square degree with $S_{350\,\mu{\rm m}} \ga 80$ mJy and
SFR $\la 200\,M_\odot$ yr$^{-1}$. To find out such galaxies extensive
follow-up observations enabling solid determinations of the gravitational
amplification are necessary.

\section{Discussion}\label{sect:disc}

\subsection{Late-type lenses}

The analysis described so far refers to spheroidal lenses. In the case of
spiral lenses, there are some significant differences. First, the lensing
cross-section depends rather strongly on the central surface mass density,
which is substantially lower in disk compared to spheroidal galaxies. For
example, let us consider a spiral galaxy within a halo as massive as that of
our reference early-type lens ($M_{\rm H}=10^{13}\, M_{\odot}$), and with the
same stellar to DM mass ratio $30$ (this ratio is found to be essentially
independent of the Hubble type for massive galaxies; see Mandelbaum et al.
2006; Shankar et al. 2006). Using the average relation by Tonini et al.
(2006b) for the corresponding stellar mass, $M_\star\approx 3\times 10^{11}\,
M_\odot$, we find a scale radius of the exponential disk profile
[$\Sigma(r)=M_\star/(2\pi\,R_D^2\, e^{-r/R_D})$]  $R_D\approx 7$ kpc. The
interstellar medium adds little: its mass is a small fraction of the stellar
mass and it is more diffuse, with a scale radius typically of $3\, R_D$.

In the radial range $r\la 10^{-1.5}\, R_{\rm H}$ (with $R_{\rm H}\approx
200\,$kpc) relevant for strong lensing the surface density of the disk is as
flat as that of the DM, and is a factor $4$ smaller. As a consequence, the
lensing properties of late-type galaxies are quite close to those of a pure
NFW configuration. For comparison, the stellar surface density of an
early-type galaxy with the same halo mass increases steeply inwards (cf.
Fig.~\ref{fig:surface_dens}); it is a factor $\approx 3$ smaller than the
DM's at $r\la 10^{-1.5}\, R_{\rm H}$ but a factor $5$ larger at $r\la
10^{-2.5}\, R_{\rm H}$. After Fig.~\ref{fig:cross_sect} this implies that the
cross sections (and relatedly the number) of massive spiral lens yielding
amplifications larger than $10$ is lower than that of an early-type with the
same mass by a factor greater than $3$.

Moreover spheroids, even though less numerous, are, on average, substantially
more massive (galaxies with $M_{\rm H}$ as large as $10^{13}\, M_{\odot}$ are
much more frequently spheroidals than spirals) and contain the major share of
stellar mass (Baldry et al. 2004). Furthermore, disk galaxies have generally
younger stellar populations than spheroidal galaxies (Bernardi et al. 2010;
their Fig. 10), indicative of a formation redshift $\la 1$; thus they are
likely increasingly rarer than spheroidal galaxies at substantial redshifts.
This explains why the contribution of spiral galaxies to the lensed counts is
subdominant; preliminary evidences of such an expectation come from lens
searches with different selection criteria (e.g., Auger et al. 2009;
Gonzalez-Nuevo et al. 2012).

\subsection{Effect of ellipticity}

Realistic lens models include some ellipticity. In fact, accounting for
ellipticity is essential in order to successfully reproduce image numbers,
image positions and extended lensed images, in most of the observed lensing
systems. As an example, axially symmetric lenses cannot produce an Einstein
cross because the tangential caustic is collapsed to a point. Therefore no
more than two (in the case of a singularity at the center like in the SIS
model) or three (in the case of a finite core density like in the NFW or in
the SISSA model) images can form. But when ellipticity is non-null, the
tangential caustic has a finite shape and once the source encompasses it a
new pair of images can form (see Fig.~\ref{fig:ellipticity}).

A direct comparison between the spherical and the ellipsoidal case can be
most easily done considering a Singular Isothermal Ellipsoid (SIE hereafter).
As we are mostly interested on the lensing statistics we focus on the effect
of  ellipticity on the  cross-section for total magnifications
$\sigma(\mu_{\rm tot})$. We consider the case of a SIE with ellipticity
$\mathrm{e}=0.2$ and of a SIE with ellipticity $\mathrm{e}=0.4$, and
assume a lens at redshift $z_\ell=0.7$, with velocity dispersion $\sigma=350$
km s$^{-1}$, and a source redshift $z_s=2.0$. The exact choice of these
values is irrelevant as one can always work in normalized units, using as a
reference scale the value of the Einstein radius for a SIS with the same
parameters (in this case $\theta_E\approx 1.85^{\prime\prime}$). We use the
public code
\textsl{GLAFIC}\footnote{http://www.slac.stanford.edu/$\sim$oguri/glafic/}, a
software developed for studying strong gravitational lenses (Oguri 2010), to
solve the lens equation for the SIE model and to estimate the total
amplification as a function of the source position in the source
plane\footnote{More precisely we have used the 'mock1' command to randomly
populate the source plane with $500000$ sources, to solve the lens equation
and get the amplifications for the individual images for each source
position. We have then constructed a map of the total amplifications by
grouping the source positions into pixels of $0.01^{\prime\prime}$ in size.
We have finally used such a map to derive the cross section $\sigma(\mu_{\rm
tot})$.}.

On the top panels of of Fig.~\ref{fig:ellipticity} we show the map of the
total amplification as a function of the source position from the center of
the lens for the SIE model with $\mathrm{e}=0.2$  (central panel) and for the
SIE model with $\mathrm{e}=0.4$ (right panel). Contours of equal
amplifications ranging from $2$ to $10$ are shown in orange. The red curves
mark the caustics. For comparison, the case of a SIS model with the same lens
and source parameters is shown on the top left panel of the same figure.

From these images the cross-section $\sigma(\mu_{\rm tot})$ is easily derived
and the results are presented in the bottom panel of
Fig.~\ref{fig:ellipticity}. We see that for amplifications below $\sim 3-4$,
the SIE model is almost indistinguishable from the SIS model in terms of
cross-section. The effect of ellipticity is a `squeezing' of the equal
amplification contours along the major axis of the lens (oriented
North-South) while leaving the enclosed area almost unaffected. On
the other hand, for amplifications close to $\mu_{\rm tot}\sim 8$ for
$\mathrm{e}=0.2$, and to $\mu_{\rm tot}\sim 4$ for $\mathrm{e}=0.4$, the
cross-section of the SIE model deviates appreciably from the regular circular
shape observed for the SIS model, as the source is now close to the inner
cross-like caustic, and becomes correspondingly smaller (by $\sim 30\%$). The
situation is reversed for higher amplifications as the cross-section for the
SIS model shrinks to a point  while that of the SIE model converges to the
finite inner caustic.  In this case the SIE model yields a cross-section that
is $\sim 50\%$ higher than that given by the SIS model and approaches
the SIS limit asymptotically for $\mu_{\rm tot}\gg 100$. The exact value of
the amplification at which the transition of the cross-section from the
`sub-SIS' to the `super-SIS' regime occurs depends on the adopted value of
the ellipticity (it is around $\mu_{\rm tot}\sim 20$ for $\mathrm{e}=0.2$ and
close to $\mu_{\rm tot}\sim 8$ for $\mathrm{e}=0.4$). In fact, as the
ellipticity is increased, the inner caustic becomes more extended and the
regions of low amplifications in the source plane are consequently more
affected.

In conclusion, compared to the case of a spherical lens, the effect
of ellipticity is to slightly decrease (by about $30\%$) the cross-section in
a range of amplifications $\mu_{\rm tot}\sim 4-20$ (the exact interval
depending on the value of the ellipticity) and to increase the cross-section
(up to $50\%$) for higher amplifications.

\bigskip\bigskip

\subsection{Evolution of the mass density profiles}\label{sect:evolution}

The mass density profile may evolve during the galaxy lifetime under the
action of several processes. For example, in the early stages of galaxy
formation when gas and stars condense toward the inner regions of the system,
`halo contraction' may lead to a steepening of the initial, NFW-like DM
profile (see Blumenthal et al. 1986; Mo et al. 1998; Gnedin et al. 2004). The
strength of the effect is widely debated, but recent numerical experiments
(see Abadi et al. 2010; Pedrosa et al. 2010; see also Gnedin et al. 2012)
suggest that the classic treatments based on adiabatic invariants are likely
to be extreme, and that actually in the inner regions the contraction may be
inefficient and the density shape hardly modified. We have checked that the
projected surface density of an overall configuration constituted by a
S\'ersic profile for the baryons and a contracted NFW profile for the DM is
still well approximated in the radial range $-2\la \log (s/R_{\rm H})\la -1$
by a power-law shape, with index steeper than the basic SISSA model.
Specifically, for a lens galaxy at $z_\ell=0.7$ with halo mass $M_{\rm
H}=10^{13}\, M_\odot$ and fiducial parameters of the mass distribution, we
find that the slope $\eta\approx 0.823$ of the basic SISSA model is steepened
to the value $0.958$ according to the prescription for halo contraction by
Blumenthal et al. (1986), to $0.920$ according to that by Gnedin et al.
(2004), and to $0.877$ according to that by Abadi et al. (2010).

On the other hand, a flattening of the mass density profile may be caused by
transfer of energy and/or angular momentum from (baryonic and DM) clumps to
the DM field during the galaxy formation process (e.g., El-Zant et al. 2001;
Tonini et al. 2006a). Moreover, at the formation of a spheroid, central
starbursts and accretion onto a supermassive black hole may easily discharge
enough energy with sufficient coupling to blow most of the gaseous baryonic
mass within the star-forming region out of the inner gravitational well. This
will cause an expansion (`puffing up') of the stellar and of the DM
distributions (see Fan et al. 2010; Ragone-Figueroa et al. 2012), so as to
flatten the inner profiles.

A reliable assessment of these competing processes is still lacking,
and beyond the scope of the present paper. Anyway, it is interesting to point
out that, since early-type galaxies generally do not show signs of
dissipation at $z\la 1$, the steepening of the density profiles due to halo
contraction should show up at higher $z$, though the trend may be partly
offset by other processes such as the energy transfer or the puffing up
mentioned above.

Recent results by Ruff et al. (2011) and Bolton et al. (2012) show
preliminary evidence for a mild evolution in the opposite direction, i.e.,
towards steeper mass profiles at later cosmic times. In fact, the average
density slopes steepen from values $\eta\approx 2$ at $z\approx 0.6$ toward
$\eta\approx 2.2$ at $z\approx 0.2$. The significativity of the detection is
still under debate, since the trend can be partly explained in terms of
variations in the lensing measurement aperture with redshift, which favours
the sampling of inner, steeper portions of the mass distribution at
decreasing $z$. In any case, we stress that around $z_\ell\approx 0.6$ where
the lens redshift distribution peaks (see Fig.~10), the measured average
density slope $\eta_V\approx 2$ (corresponding to a surface density slope
$\eta\approx 0.8$, see \S~2.2) is in excellent agreement with the SISSA model
outcomes. In addition, even if the trend toward steeper slopes $\eta\ga 2.2$
(i.e., $\eta_V\ga 1$) at $z\approx 0.2$ will be confirmed, our lensing
analysis based on power-law representations of the surface density would
still apply. The overall effect on the amplification distribution would be
small since the lens redshift distribution is steeply declining for $z\la
0.5$, although in the phenomenology of individual lensing systems (e.g.,
Einstein ring size; presence/absence of inner critical curve) the local
lenses would tend to behave as ideal SIS more than high-redshift ones.

\subsection{Super-massive black holes in the lens centers}

Super-massive black holes (BHs) are ubiquitous in the nuclei of early-type
galaxies (see Ferrarese \& Ford 2005 for an exhaustive review). What is the
effect of a supermassive BH in the nucleus of a lens galaxy? For an AGN in
the source the analysis of point source lensing (\S\,\ref{sect:cross_sec} and
\ref{sect:solutions}) applies. A thorough discussion will be presented in a
subsequent paper.

By itself, an isolated point mass $M_\bullet$ features a surface
density $\Sigma(\theta)\propto M_\bullet\, \delta_D(\theta)/\theta$ in terms
of the Dirac delta function $\delta_D(\theta)$, which yields a deflection
profile $\alpha(\theta)\propto M_\bullet/\theta$; in our formalism based on
Eqs.~(9) and (13) this configuration corresponds, approximately for
$\Sigma\propto \theta^{-\eta}$ and exactly for $\alpha\propto
\theta^{1-\eta}$, to the limiting powerlaw index $\eta\rightarrow 2$. Then
after Eq.~(18) it is seen that two images are produced independently of the
impact parameter (actually there is a third image but it has zero
magnification), and no inner critical curve is present [cf.
\S~\ref{sect:solutions}]. Normalizing angles in units of $\theta_E\rightarrow
[M_\bullet\,(1+z_\ell)^2/\pi\,D_\ell^2\,\Sigma_c]^{1/2}$, the locations of
the images are $\bar\theta=(\bar\beta\pm \sqrt{4+\bar\beta^2})/2$ and their
magnifications are $\mu_\pm=\pm\,(\bar\beta/\sqrt{4+\bar\beta^2}+
\sqrt{1+4/\bar\beta^2}\pm 2)/4$ so that the total magnification reads
$\mu=(2+\bar\beta^2)/\bar\beta\sqrt{4+\bar\beta^2}$. When the source
approaches the optical axis at $\beta\rightarrow 0$, the total magnification
diverges as $\beta^{-1}$ and the image positions tend to $\theta_E$ (for
details, see  Kochanek 2006).

Now we turn to discuss the effect of a supermassive BH at the center of a
galactic structure. For sake of definiteness, let us consider a lens galaxy
with $M_{\rm H}=10^{13}\,M_\odot$ and a mass density profile described by our
fiducial SISSA model with concentration $c=5$, S\'ersic index $n=4$, and dark
to baryon mass ratio $M_{\rm H}/M_\star=30$. To this configuration we add a
central supermassive BH, with standard BH to stellar mass ratio
$M_\bullet/M_\star\approx 2.5\times 10^{-3}$.

The overall surface density is illustrated in Fig.~\ref{fig:BH1};
here, to avoid dealing with the central singularity associated to the
BH, we rendered its surface density with the powerlaw $\Sigma(\theta)\propto
\theta^{-\eta}$ in the limit $\eta\rightarrow 2$ (see above). It is seen
that the BH dominates only at radii $s/R_{\rm H}\la 10^{-3}$.

In Fig.~\ref{fig:BH2} we show the solutions of the lensing equation, in terms
of the $\beta-\theta$ and $\beta-\mu$ relations. These figures should be
contrasted with Fig.~\ref{fig:surface_dens} and \ref{fig:solutions} that
represent the same lens configuration without central supermassive BH. All in
all, the presence of the BH produces three effects: first, the radius of the
Einstein ring is increased, although only slightly given the small BH
contribution to the enclosed mass; second, the inner critical curve is
erased, since the steep surface density of the point-mass lens dominates the
overall behavior for $\theta\rightarrow 0$; third, the central demagnified
image can be accompanied by a second detectable central image when the BH
mass falls in the range $10^{7.5}-10^{8.5}\,M_\odot$ (Rusin, Keeton \& Winn
2005).

In Fig.~\ref{fig:BH3} we show that the central supermassive BH has only a
minor effect on the differential amplification distribution. In this example,
we have adopted two values of the BH to stellar mass ratio, and specifically
a standard one $M_\bullet/M_\star=2.5\times 10^{-3}$ and a high one
$2.5\times 10^{-2}$ as measured recently in two giant ellipticals by
McConnell et al. (2011). We have also checked that the mass/redshift
dependent $M_\bullet-M_\star$ relationship from the galaxy formation model of
Lapi et al. (2006, 2011) yields outcomes similar to the former case.

\subsection{Effect of a galaxy cluster halo surrounding the lens galaxy}

The presence of a galaxy cluster halo around an early-type lens can strongly
affect the lensing phenomenology of individual objects. However, if
the lens system is roughly centered on the cluster halo, we find that the
effects of such events on the amplification distribution are minor. This can
be understood on considering the low abundance of cluster-sized halos in the
halo mass function, rapidly decreasing with increasing redshift, and taking
into account that the distribution of lens redshifts typically peaks at
substantial redshifts (Fig.~\ref{fig:zdistr}). For example, consider an
early-type lens with $M_{\rm H}=10^{13}\, M_{\odot}$ at the center
of a cluster halo of $10^{15}\, M_{\odot}$. In the radial range relevant for
strong lensing ($r\sim 10^{-2.5}-10^{-1}\, R_{\rm H}$ with a galactic halo
size $R_{\rm H}\approx 200$ kpc) the cluster halo contributes $\approx
2\times 10^{10}-7\times 10^9\, M_\odot$ kpc$^{-2}$ to the surface density,
i.e., an amount comparable or slightly larger than that due to the mass
within the galaxy. On the other hand, at $z\approx 0.7$ an halo of $10^{15}\
M_\odot$ is rarer than a halo of $10^{13}\,M_\odot$ by a factor $10^6$,
implying that the contribution of these events to the amplification
distribution are negligible.

On the other hand, some contribution may arise from events in which
the lens system is not centered on the cluster halo, but is in the vicinity
of a galaxy cluster/group in projection. This is because in such instances
the cluster/group induces an external shear on the lens system, that breaks
the spherical symmetry and leads to astroid caustics similar to those arising
from a non-null ellipticity. As discussed in \S~5.2, the lensing cross
section is correspondingly affected and, considering that the statistics of
such events may be higher than that of lens systems centered with the cluster
halo (though it is difficult to provide an educated estimate), the
contribution to the amplification distribution may also be non-negligible.

\section{Conclusions}\label{sect:concl}

\begin{figure}
\epsscale{0.75}
\plotone{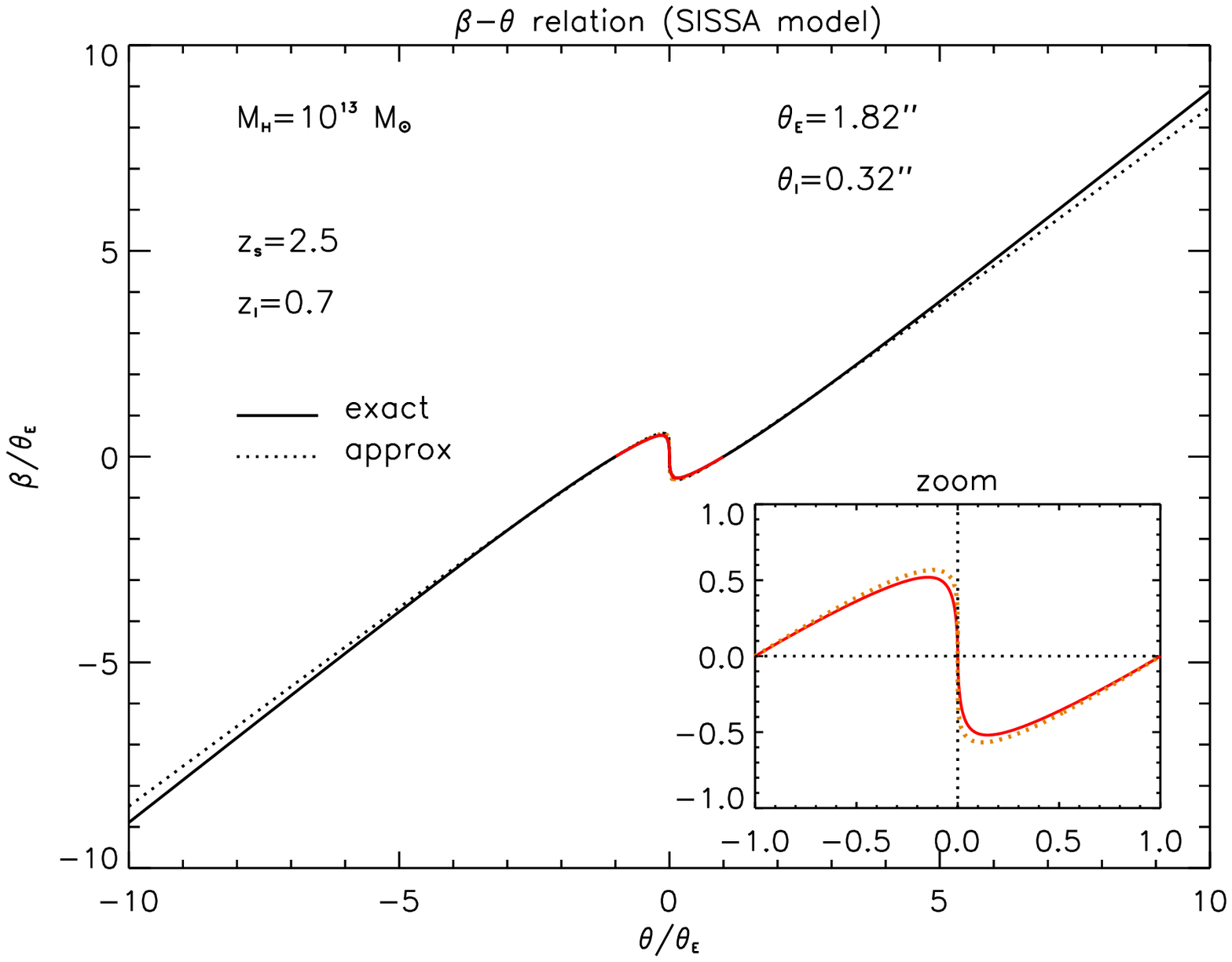}\\
\plotone{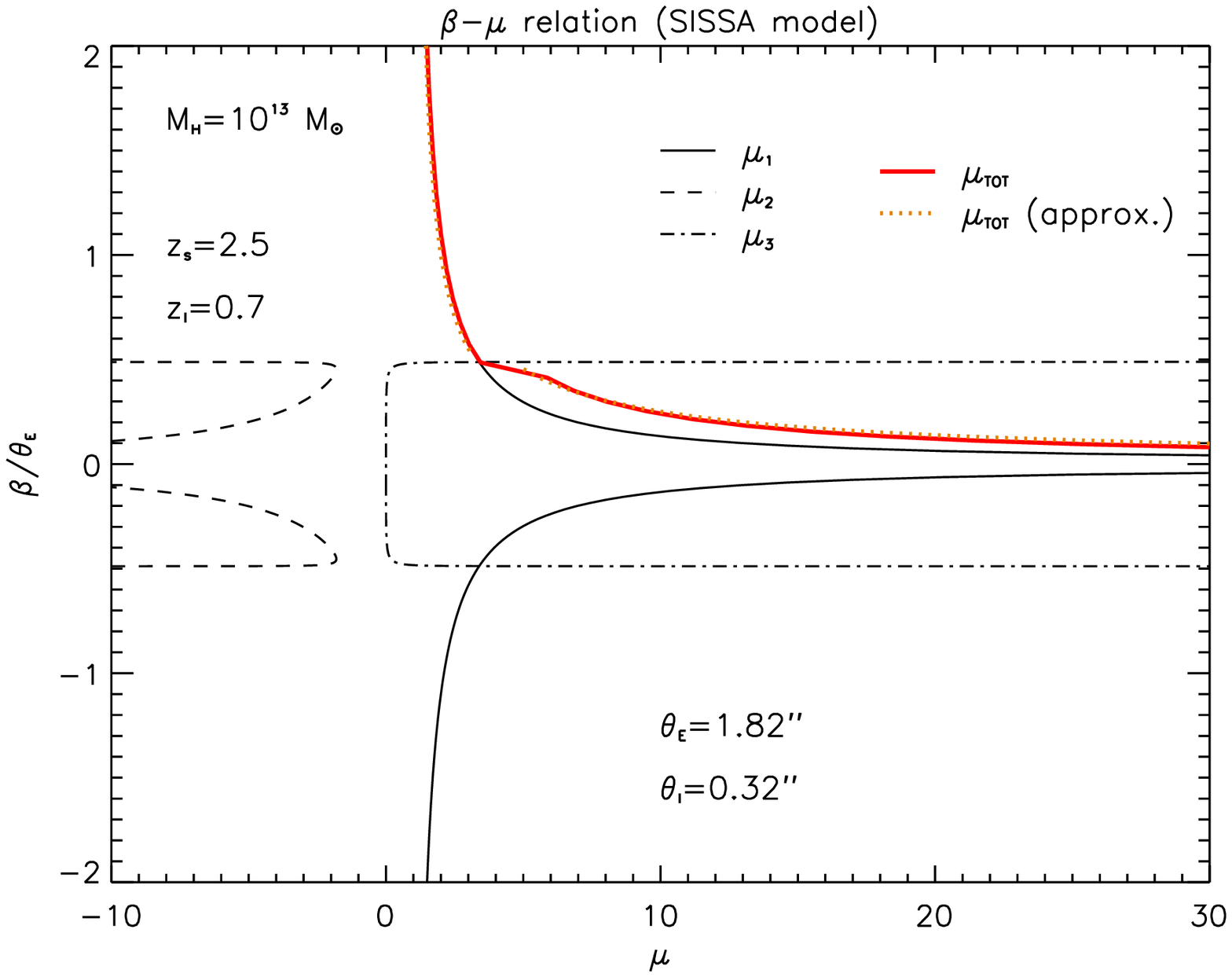}
\caption{Solutions of the lens equation for the SISSA model. \textit{Top
panel}: relation between the angular positions of the source $\beta$ and of
the images $\theta$, normalized to the position $\theta_E$ of the Einstein
ring (the red portion of the curve is zoomed in the inset). \textit{Bottom
panel}: relation between the angular position of the source $\beta$ and the
amplification $\mu_{\rm i}$ of the three images (the black solid,
dashed and dot-dashed curves refer to different images). The red solid curve
is the total magnification. In this example, the source is at $z_s=2.5$ and
the lens at $z_\ell=0.7$. The parameters of the lens mass distribution are
the same as in Fig.~\protect\ref{fig:surface_dens}. The Einstein ring is
located at $\theta_E\approx 1.82^{\prime\prime}$. Note that the SISSA and NFW
models (but not the SIS model) also feature an inner critical curve
at $\theta_I\approx 0.32^{\prime\prime}$. In both panels the orange dotted
curves show the results obtained using the approximate solutions presented in
\S\,\protect\ref{sect:solutions}; the agreement with numerical solutions is
strikingly good.}\label{fig:solutions}
\end{figure}

\begin{figure}
\center
\includegraphics[height=10cm]{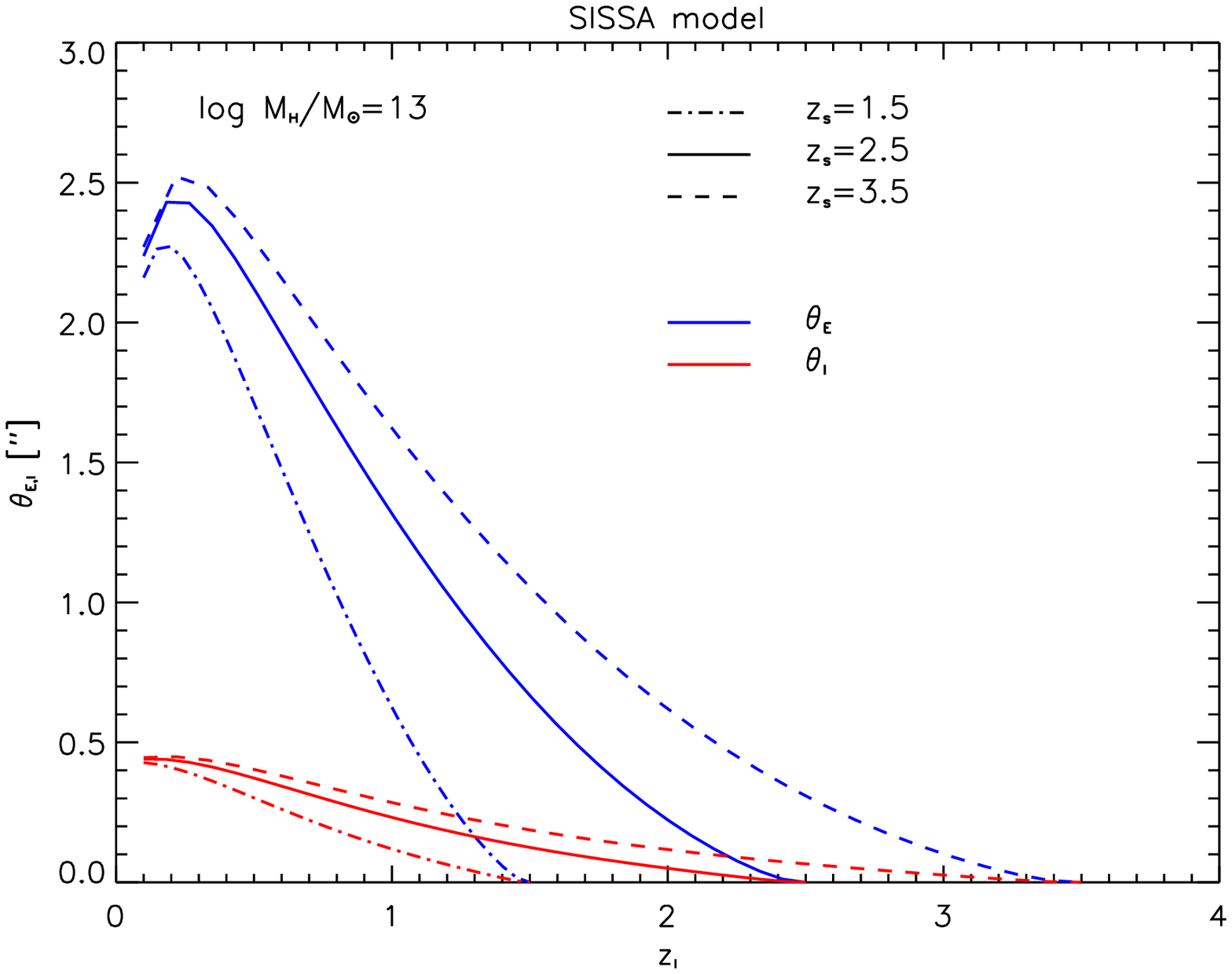}\\
\includegraphics[height=10cm]{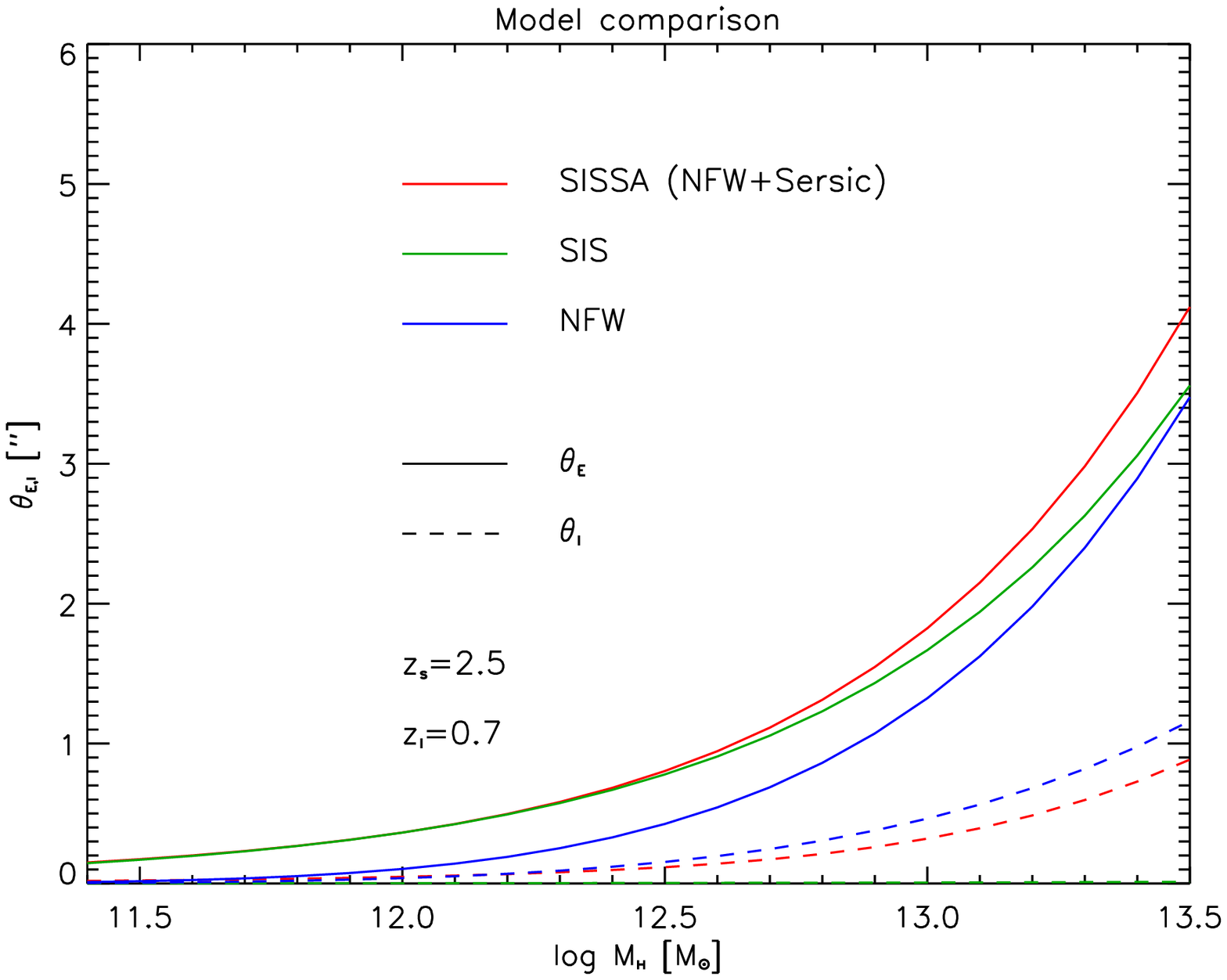}
\caption{\textit{Top panel}: angular positions of the Einstein ring
$\theta_E$ (blue lines) and of the inner critical curve $\theta_I$
(red lines) for the SISSA model, as a function of the lens redshift $z_\ell$
and for different source redshifts: $z_s=1.5$ (dot-dashed lines); $2.5$
(solid lines); and $3.5$ (dashed lines). The parameters of the lens mass
distribution are the same as in Fig.~\protect\ref{fig:surface_dens}.
\textit{Bottom panel}: angular positions of the Einstein ring $\theta_E$
(solid lines) and of the inner critical curve $\theta_I$ (dashed
lines) as a function of the lens mass, for the SISSA (red lines), NFW (blue
lines), and SIS (green lines) models. The source is at redshift $z_s=2.5$,
the lens at $z_\ell=0.7$.}\label{fig:angles}
\end{figure}

\begin{figure}
\center
\includegraphics[height=10cm]{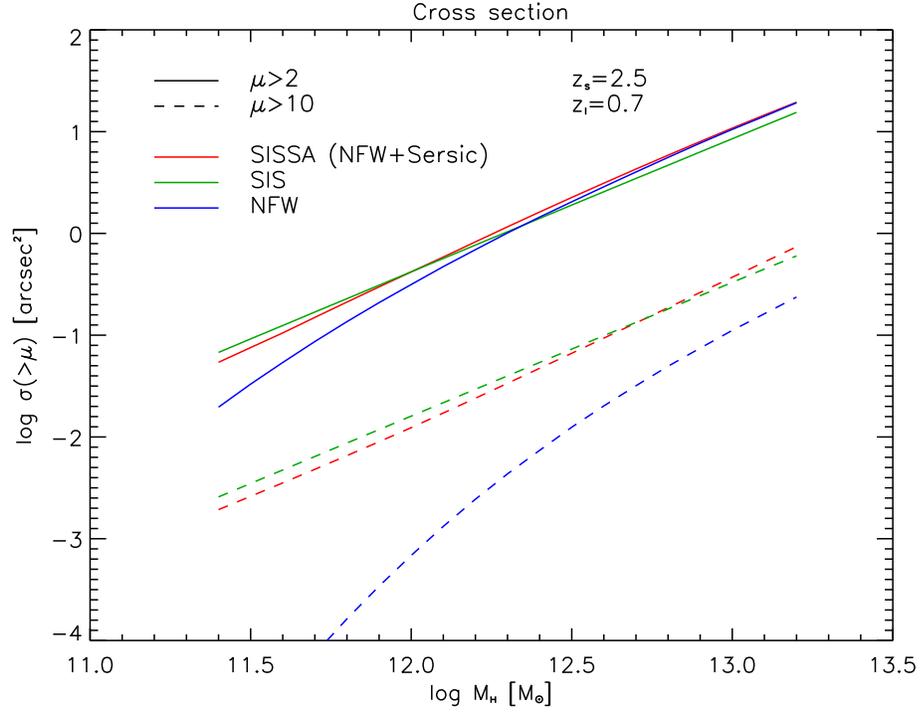}
\caption{Lens cross section as a function of the lens mass, for two different
amplification thresholds ($\mu>2$, solid lines, and $\mu >10$, dashed lines),
and for the SISSA (red lines), NFW (blue lines), and SIS (green lines)
models. The source and the lens are at the same redshifts as in
Fig.~\protect\ref{fig:angles}. }\label{fig:cross_sect}
\end{figure}

\begin{figure}
\center
\includegraphics[height=16cm]{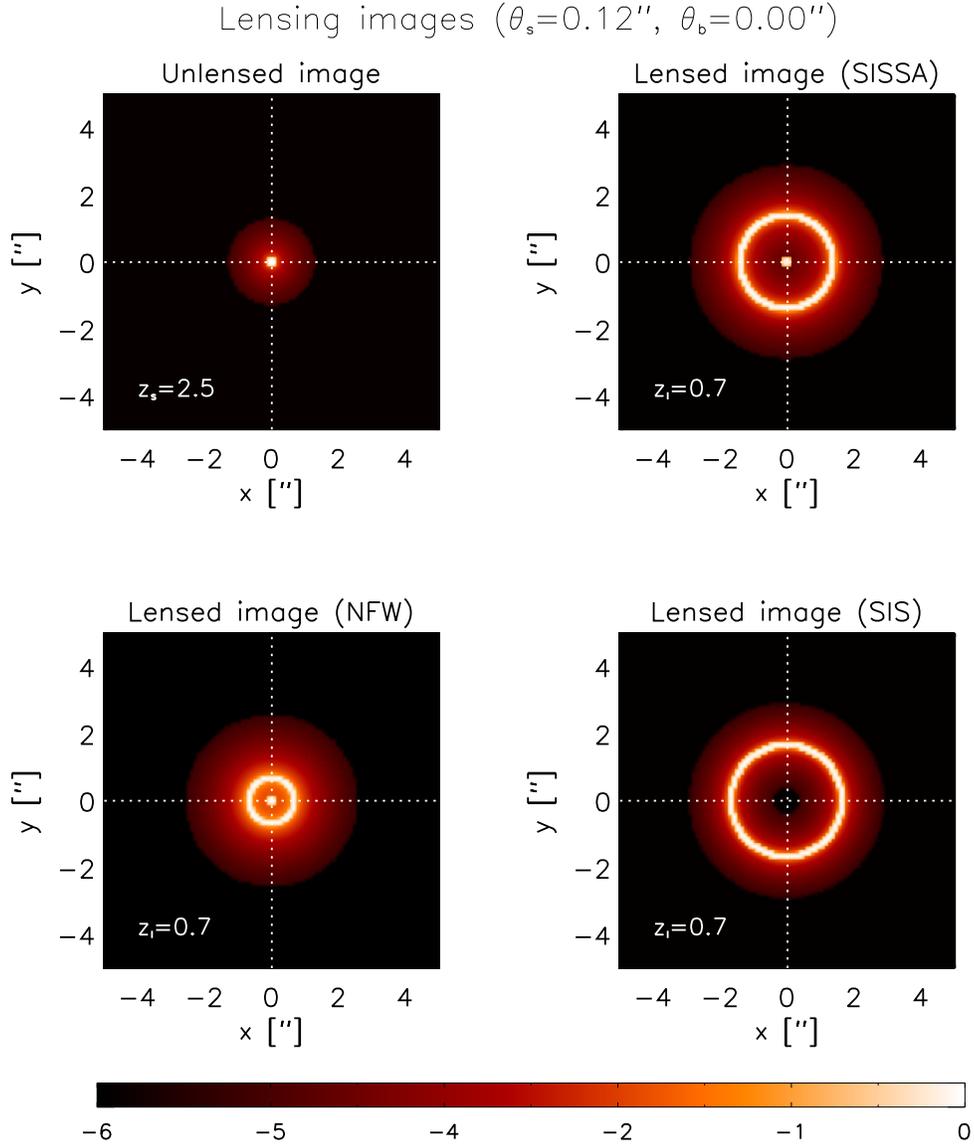}
\caption{Ray-tracing simulation of the gravitational lensing for an extended
source at $z_s=2.5$ with a S\'ersic profile ($n=4$) and effective (half
light) angular radius $\theta_s=0.12^{\prime\prime}$, and a lens at
$z_\ell=0.7$. The lens parameters are the same as in
Fig.~\protect\ref{fig:surface_dens}. The impact parameter $\theta_b$ (i.e.
the angular separation between the source and the optical axis) is null in
this example. The unlensed source is shown in the \textit{top left panel},
while the lensed image for the SISSA model is in the \textit{top right
panel}, for the NFW model in the \textit{bottom left panel}, and for the SIS
model in the \textit{bottom right panel}. In all panels the origin of
coordinates marks the position of the optical axis, while the color scale
represents in logarithmic units the surface brightness relative to the
central value of the unlensed image. Note that the central pixels in
the SISSA and NFW panels correspond to the strongly demagnified image, and
appear in the figure owing to the finite resolution of the
simulations.}\label{fig:ray_tr1}
\end{figure}

\begin{figure}
\center
\includegraphics[height=16cm]{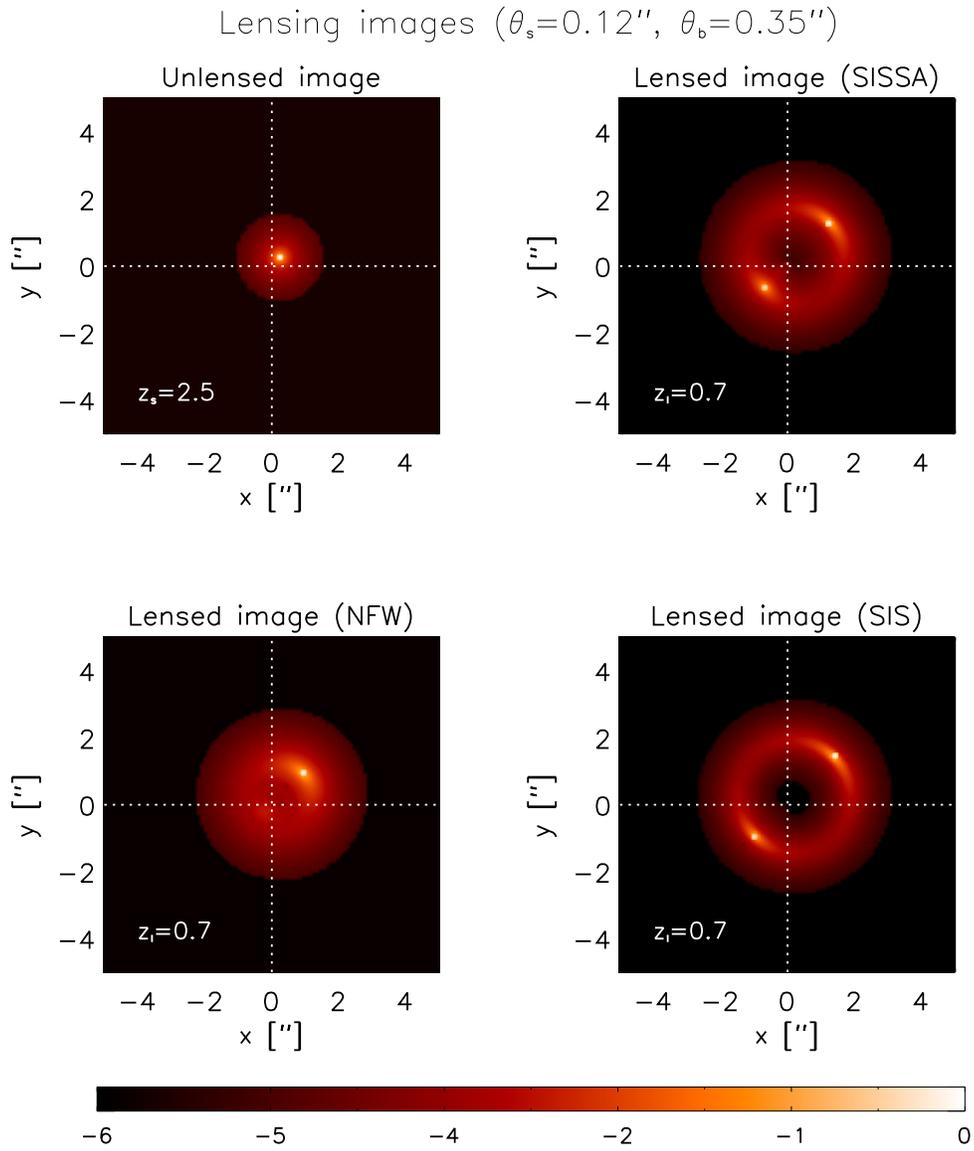}
\caption{As in the previous figure, but for an impact parameter
$\theta_b=0.35^{\prime\prime}$.}\label{fig:ray_tr2}
\end{figure}

\begin{figure}
\center
\includegraphics[height=16cm]{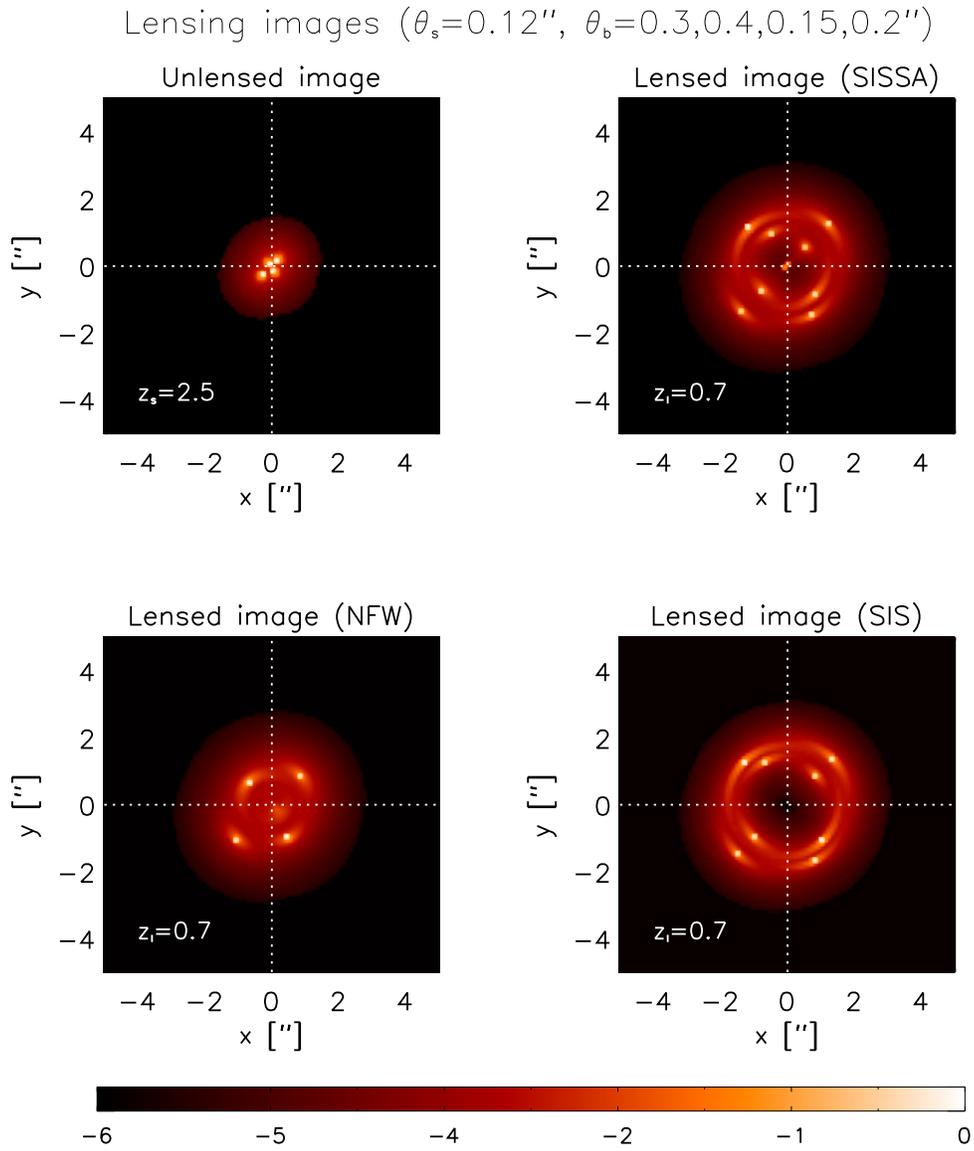}
\caption{As in the previous figures, but for four bright spots with impact
parameters $\theta_b=0.3, 0.4, 0.15, 0.2^{\prime\prime}$,
respectively.}\label{fig:ray_tr3}
\end{figure}

\begin{figure}
\center
\includegraphics[height=10cm]{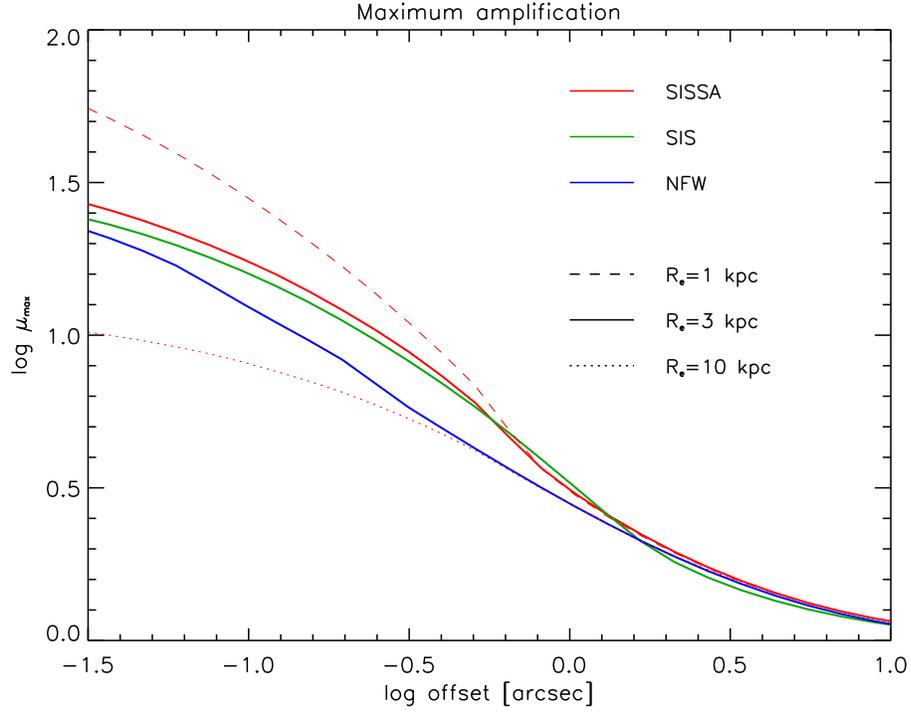}
\caption{Maximum amplification for an extended source as a function of the
offset between the center of the source and the optical axis, for the SISSA
(red line), NFW (blue), and SIS (green) models. The lens parameters and the
source redshift are the same as in Fig.~\protect\ref{fig:surface_dens}. The
source surface brightness profile is described by a S\'ersic law with $n=4$
and effective radius $R_e=3$\,kpc (solid lines). For the SISSA model we also
show the results with $R_e=1$\,kpc (dashed line) and $R_e=10$\,kpc (dotted
line).}\label{fig:max_ampl}
\end{figure}

\begin{figure}
\center
\includegraphics[height=10cm]{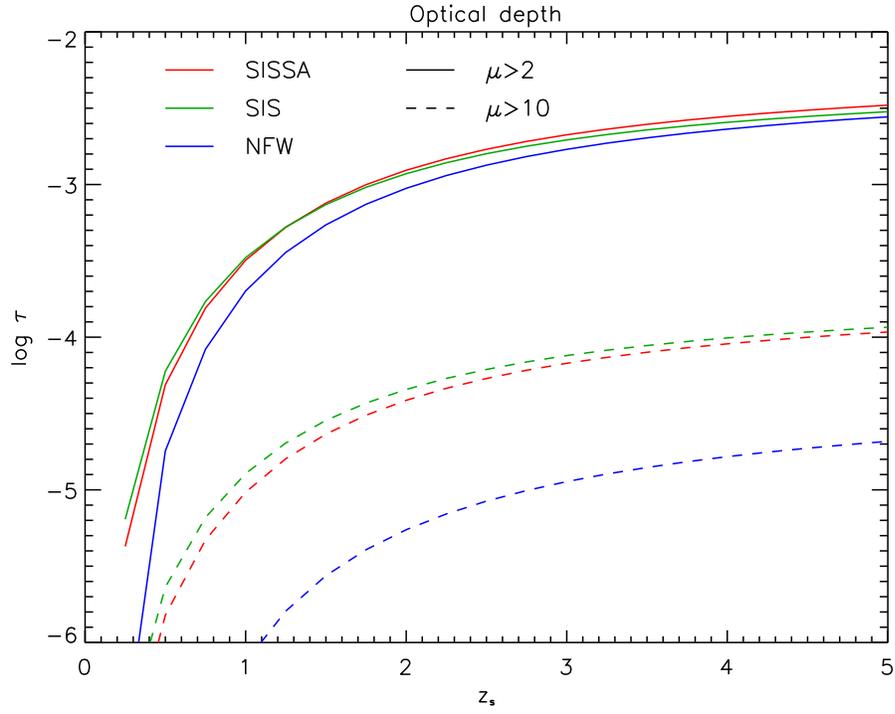}
\caption{Lensing optical depth as a function of the source redshift, for two
different amplificationx thresholds ($\mu>2$, solid lines and $\mu >10$,
dashed lines), and for the SISSA (red lines), NFW (blue lines), and SIS
(green lines) models.}\label{fig:tau}
\end{figure}

\begin{figure}
\center
\includegraphics[height=10cm]{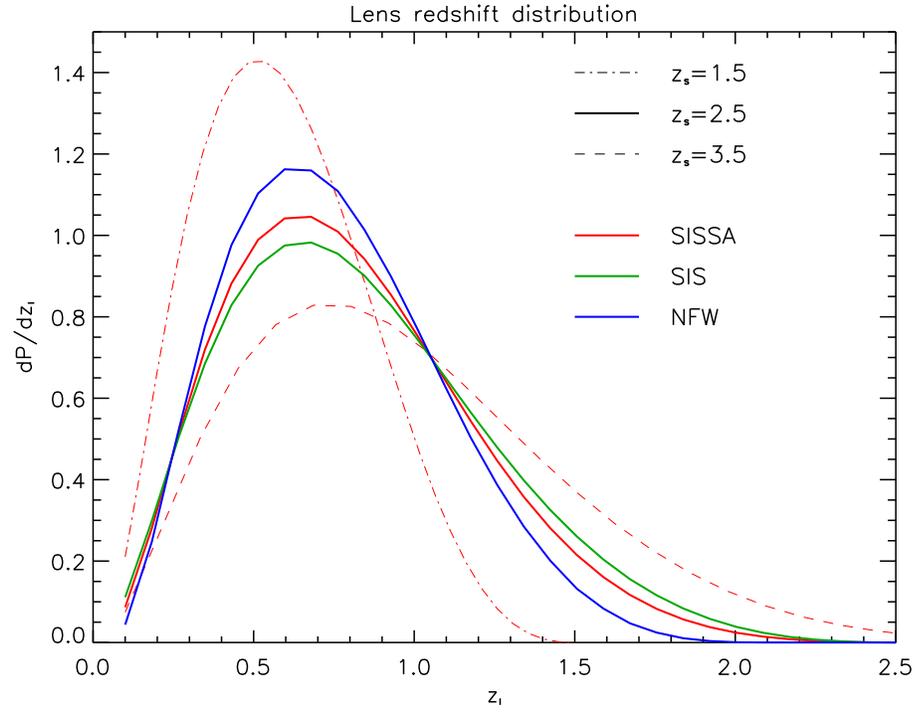}
\caption{Lens redshift distributions yielded by the SISSA (red lines), NFW
(blue line), and SIS (green line) models for a source redshift $z_s=2.5$
(solid lines). For the SISSA model we also show the results for $z_s=1.5$
(dot-dashed line) and $z_s=3.5$ (dashed line).}\label{fig:zdistr}
\end{figure}

\begin{figure}
\center
\includegraphics[height=10cm]{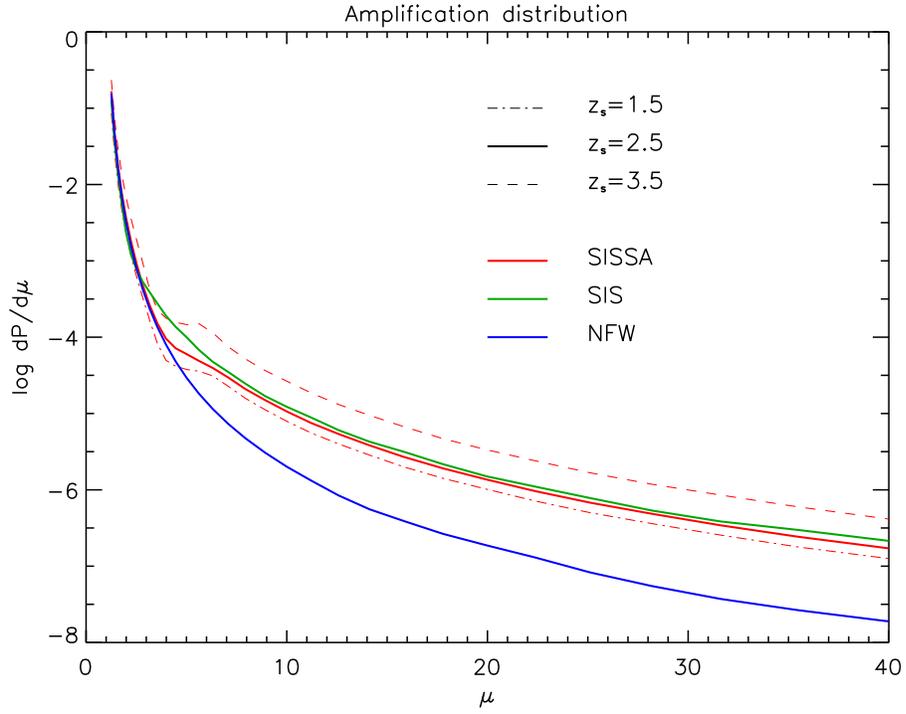}
\caption{Amplification distributions yielded by the SISSA (red lines), NFW
(blue line), and SIS (green line) models for  $z_s=2.5$ (solid lines). For
the SISSA model we also show the results for $z_s=1.5$ (dot-dashed line) and
$z_s=3.5$ (dashed line).}\label{fig:amplif_distr}
\end{figure}

\begin{figure}
\center
\includegraphics[height=16cm]{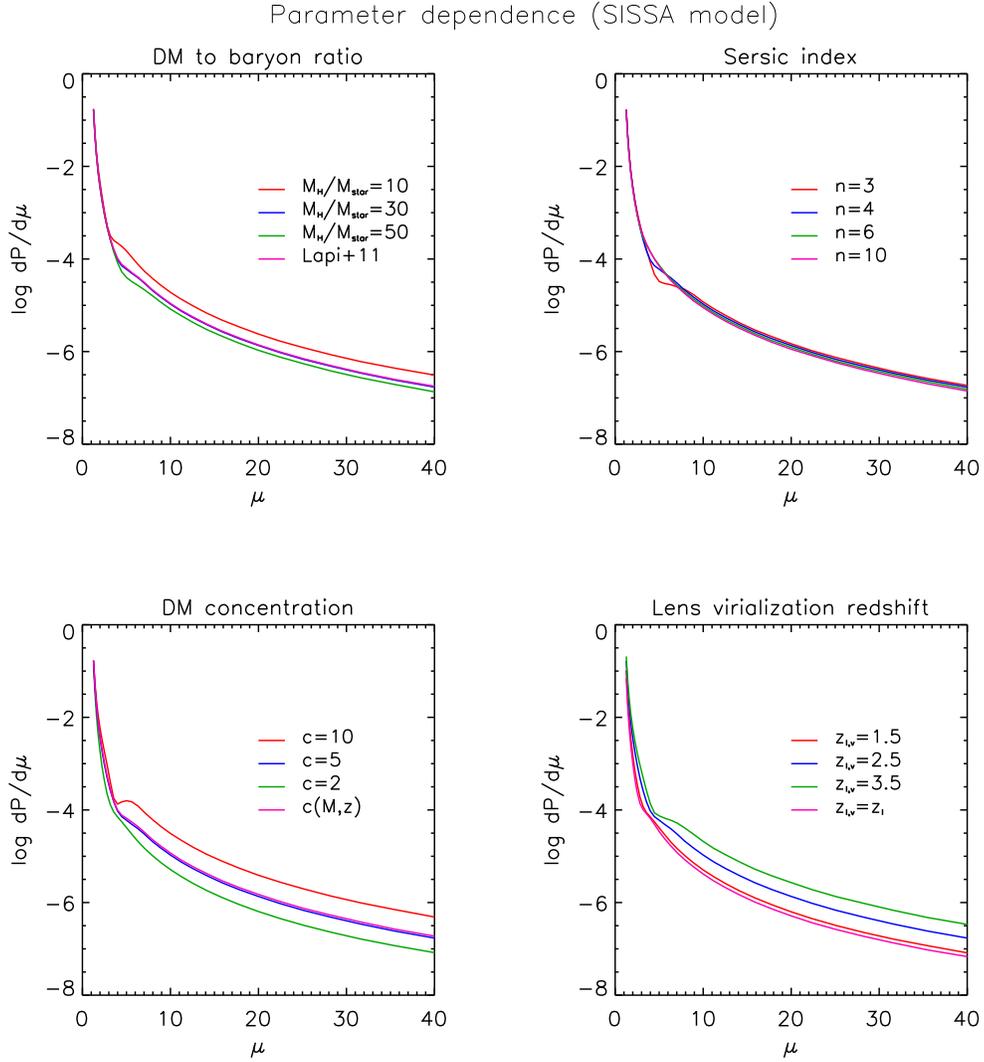}
\caption{Dependence of the amplification distribution for the SISSA model on
various parameters: DM to stellar mass ratio $M_{\rm H}/M_\star$ (\textit{top left
panel}); S\'ersic index $n$ (\textit{top right panel}); DM concentration $c$
(\textit{bottom left panel}), and lens virialization redshift $z_{\ell,v}$
(\textit{bottom right panel}).}\label{fig:parameters}
\end{figure}

\begin{figure}
\center
\includegraphics[height=10cm]{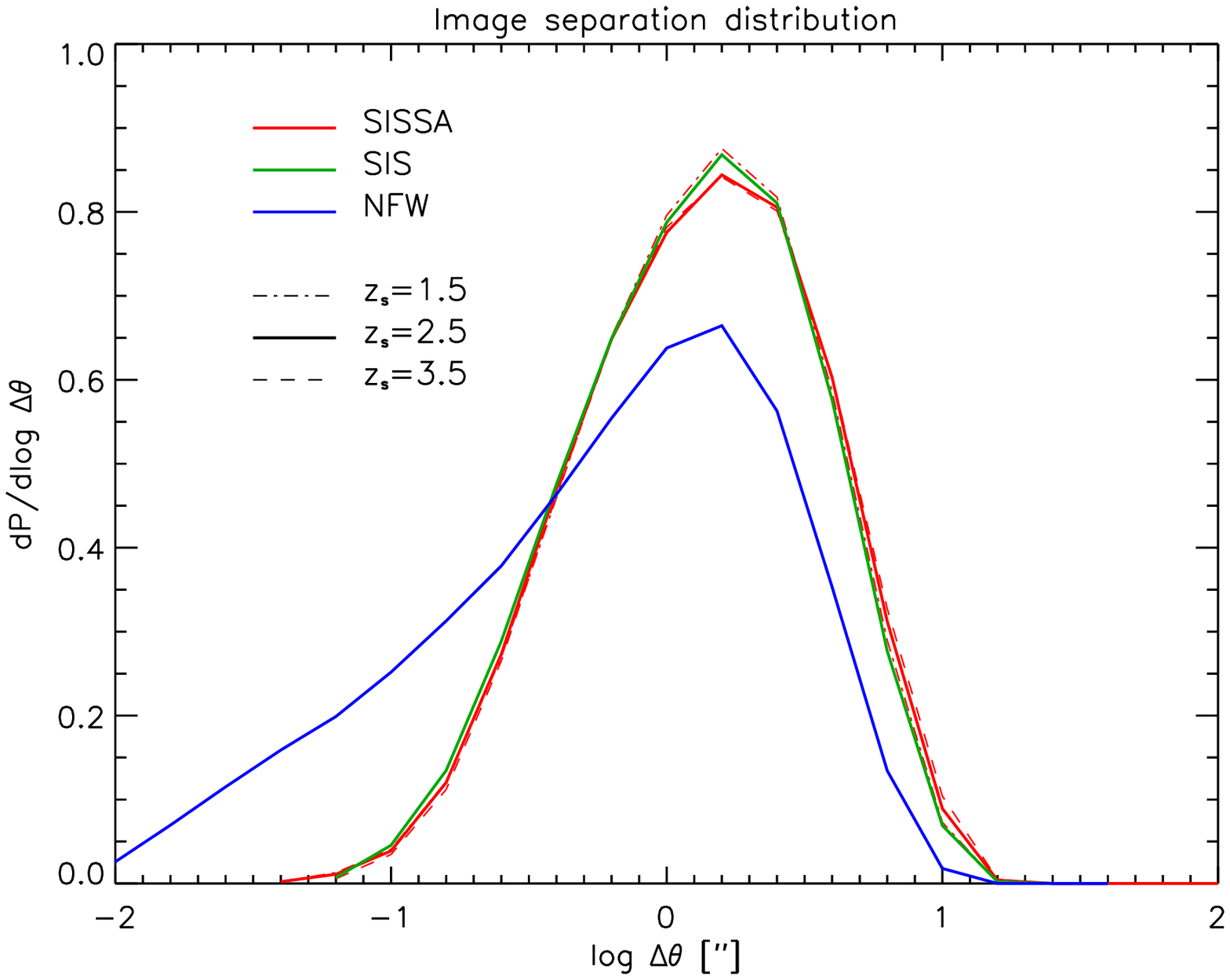}\\
\includegraphics[height=10cm]{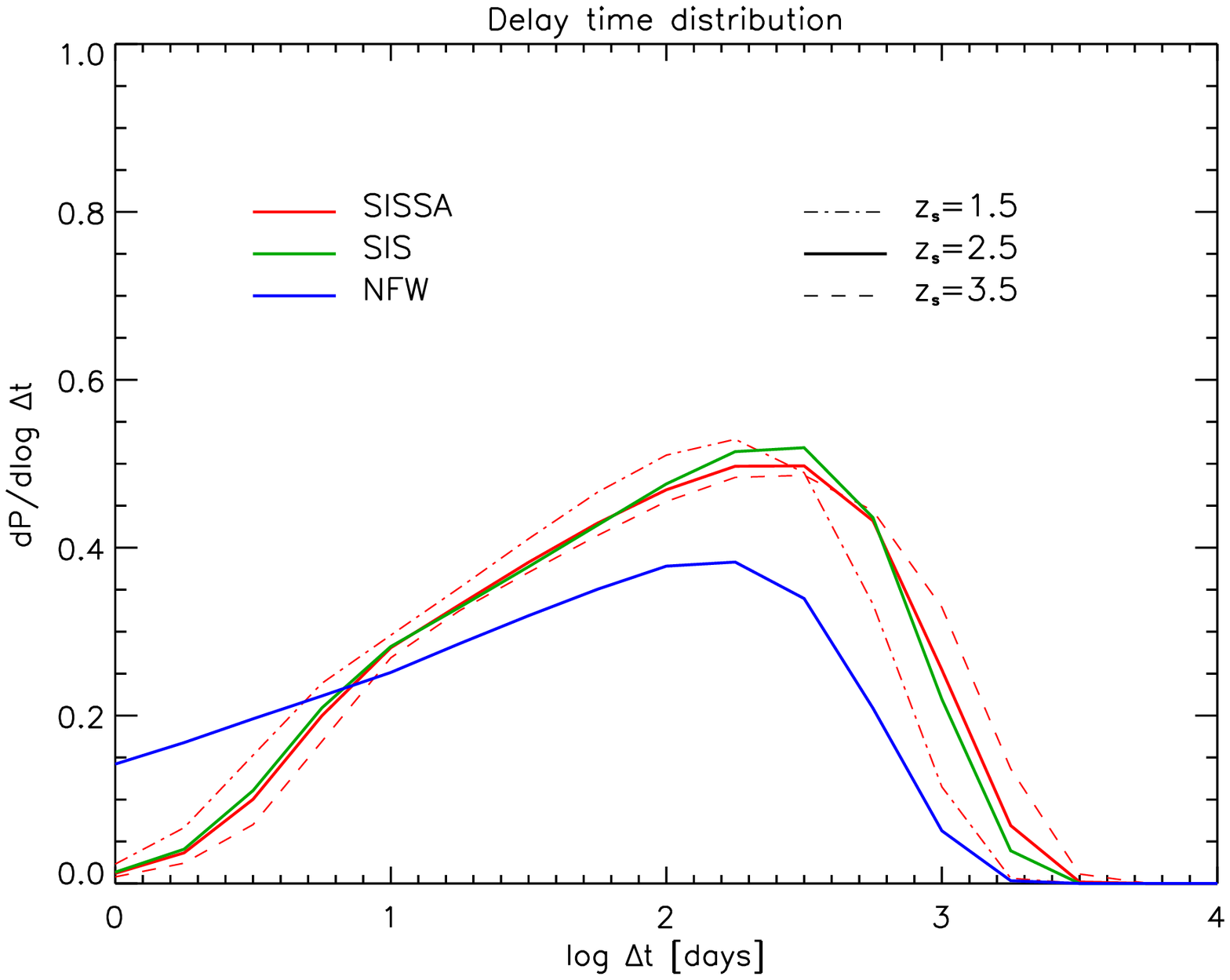}
\caption{\textit{Top panel}: distribution of the separations of the two
brightest images of a source at $z_s=2.5$ for the SISSA (red lines), NFW
(blue line), and SIS (green line) models. For the SISSA models we also show
the results for $z_s=1.5$ (dot-dashed line) and $z_s=3.5$ (dashed line).
\textit{Bottom panel}: delay time distribution; linestyles are as
above.}\label{fig:distributions}
\end{figure}

\begin{figure}
\center
\includegraphics[height=9.5cm]{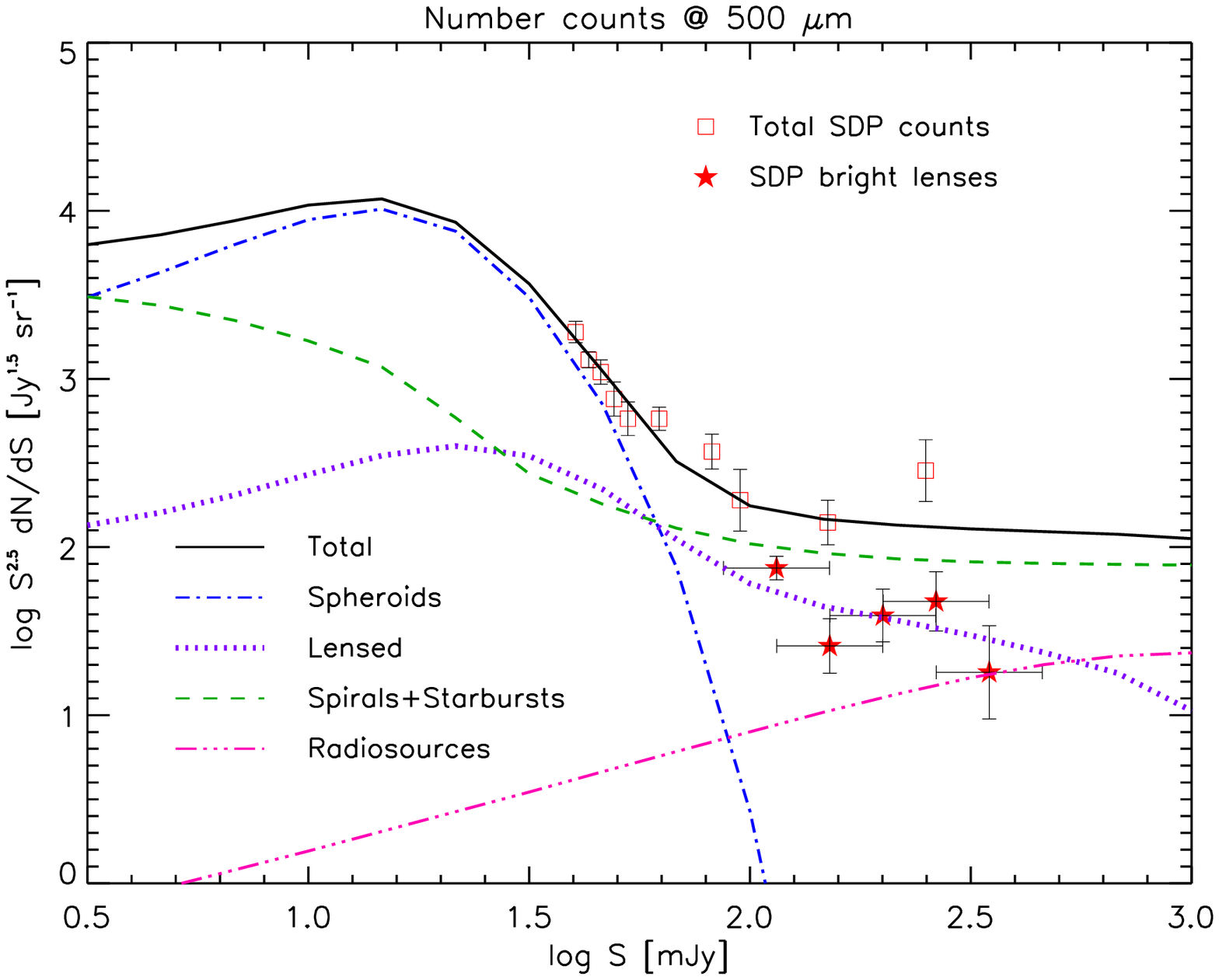}\\
\includegraphics[height=9.5cm]{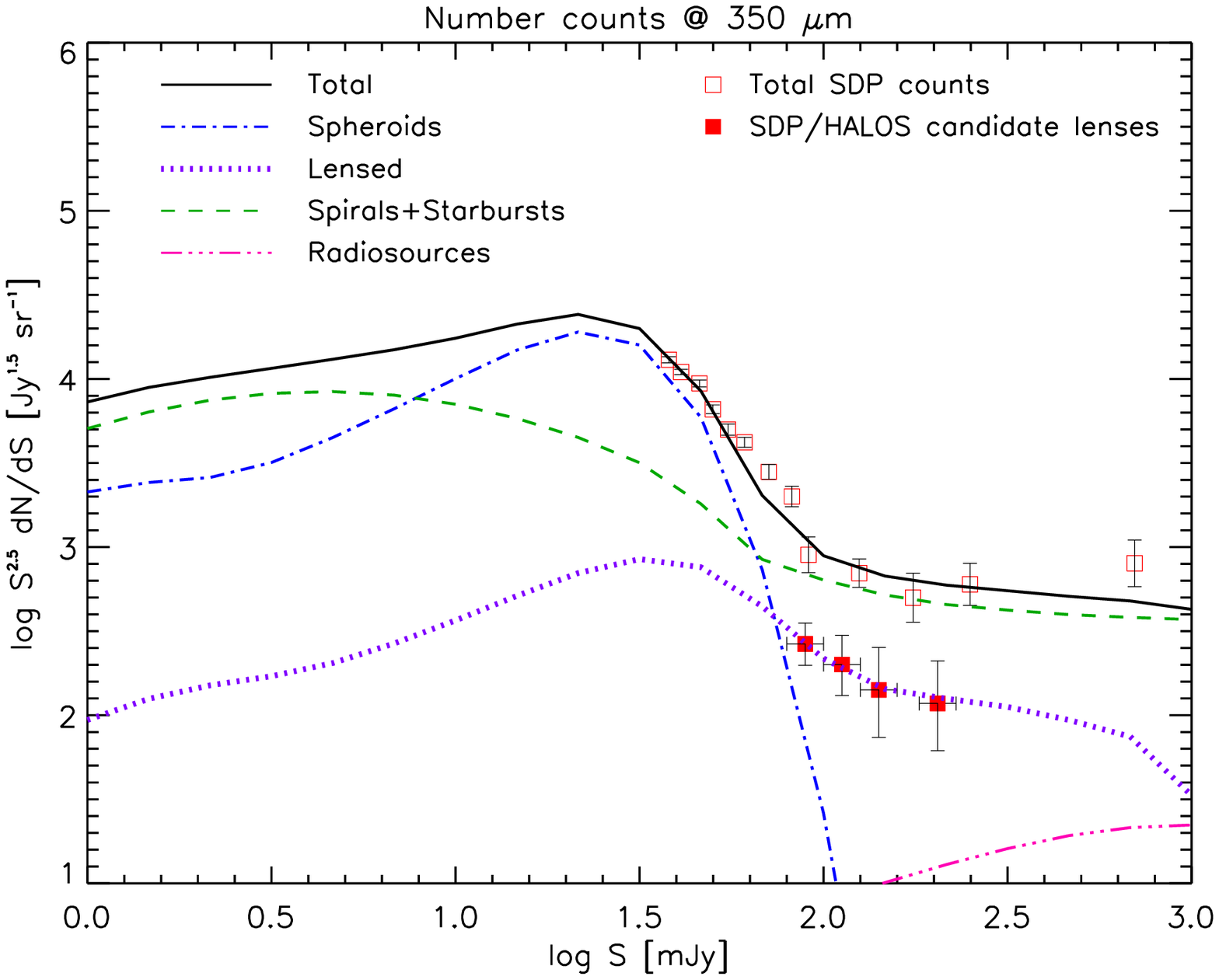}
\caption{\textit{Top panel}: Euclidean normalized counts at $500\,\mu$m. The
open squares and the filled stars represent the total \textsl{Herschel}-ATLAS
SDP counts (Clements et al. 2010) and the counts of the bright SDP lenses
spectroscopically confirmed by Negrello et al. (2010), respectively. The
solid line illustrates the total model counts comprising the contributions of
unlensed (blue dot-dashed line; Lapi et al. 2011) and strongly lensed (purple
dotted line; SISSA model from the present work) proto-spheroidal galaxies,
normal late-type plus starburst galaxies (green dashed line; Negrello et al.
2007), radio sources (magenta triple-dot-dashed line; De Zotti et al. 2005).
\textit{Bottom panel}: same but at $350\,\mu$m. Here the filled squares refer
to counts of the SDP candidate lenses selected by Gonzalez-Nuevo et al.
(2012).}\label{fig:counts}
\end{figure}

\begin{figure}
\center
\includegraphics[height=10cm]{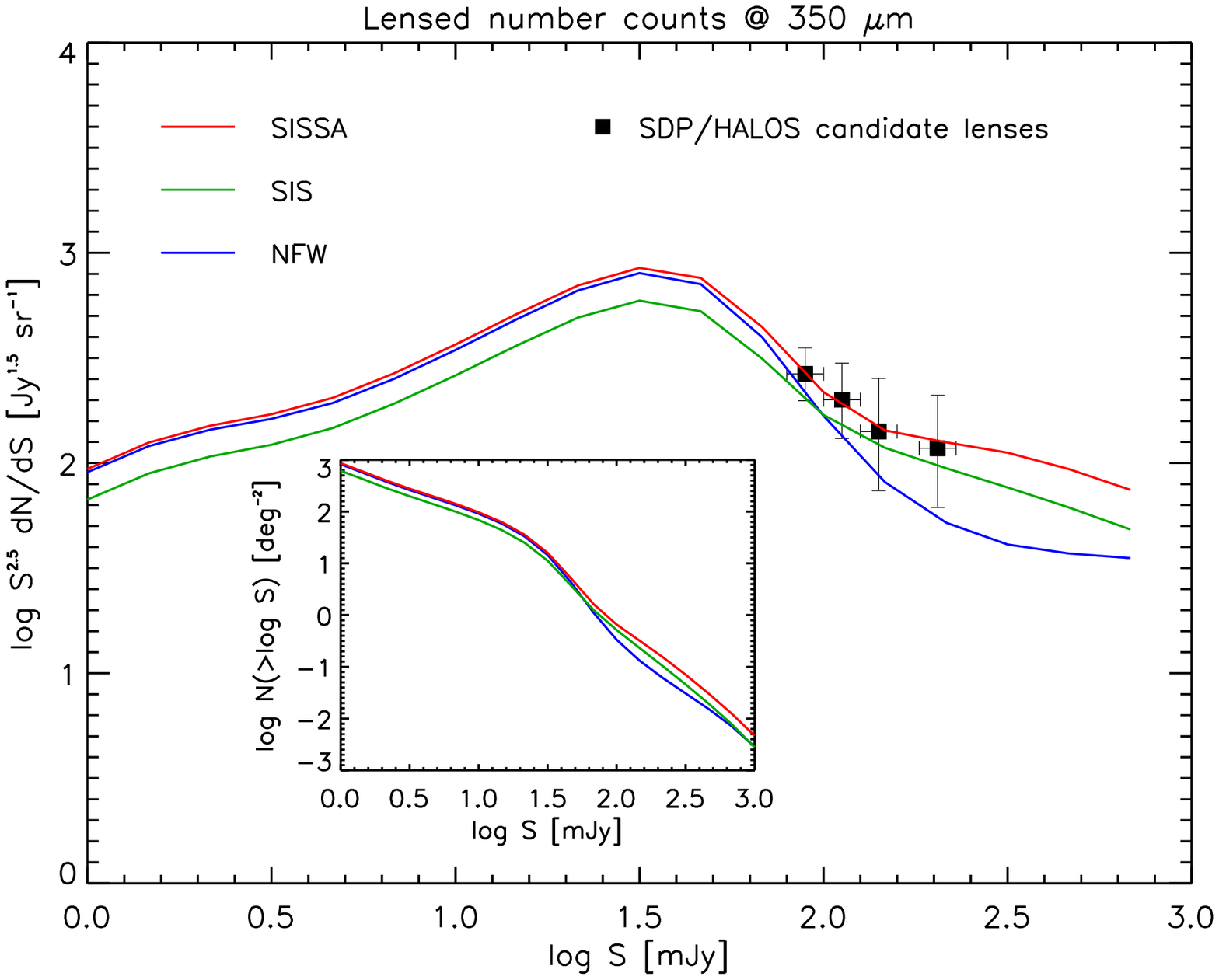}\\
\includegraphics[height=10cm]{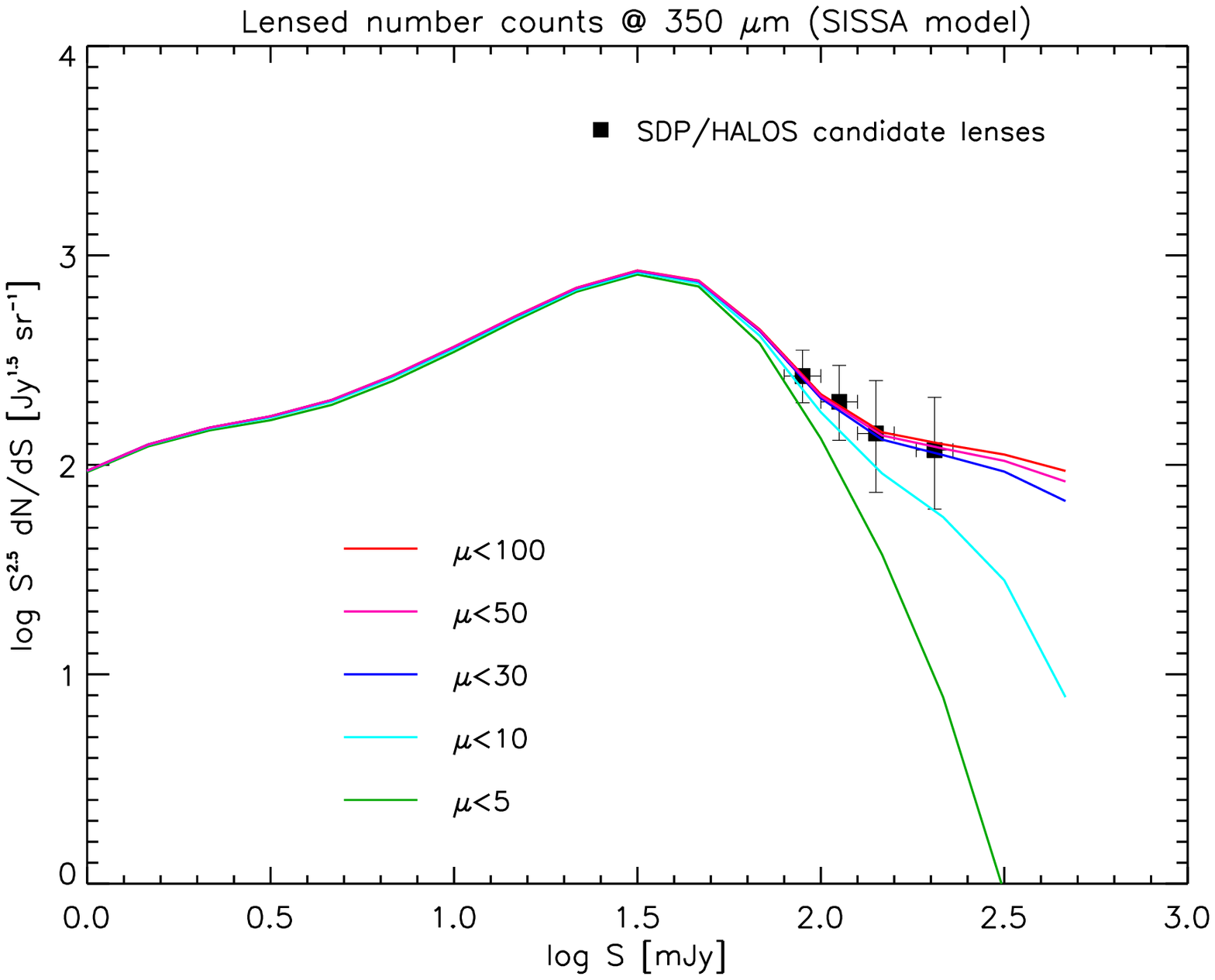}
\caption{\textit{Top panel}: Euclidean normalized counts of lensed galaxies
at $350\,\mu$m. The filled squares refer to the SDP candidate strongly lensed
galaxies selected by Gonzalez-Nuevo et al. (2012). Our model predictions are
also plotted, for the SISSA (red lines), NFW (blue lines), and SIS (green
lines) models. In the inset the corresponding integral counts are plotted.
\textit{Bottom panel}: same as above, but for the SISSA model with different
cuts in maximum amplification: $\mu_{\rm max}=5$, 10, 30, 50, and 100 (green,
cyan, blue, magenta, and red line, respectively). The line with $\mu_{\rm
max}=100$ is indistinguishable from that without amplification
cut.}\label{fig:zoom_counts350}
\end{figure}

\begin{figure}
\center
\includegraphics[height=10cm]{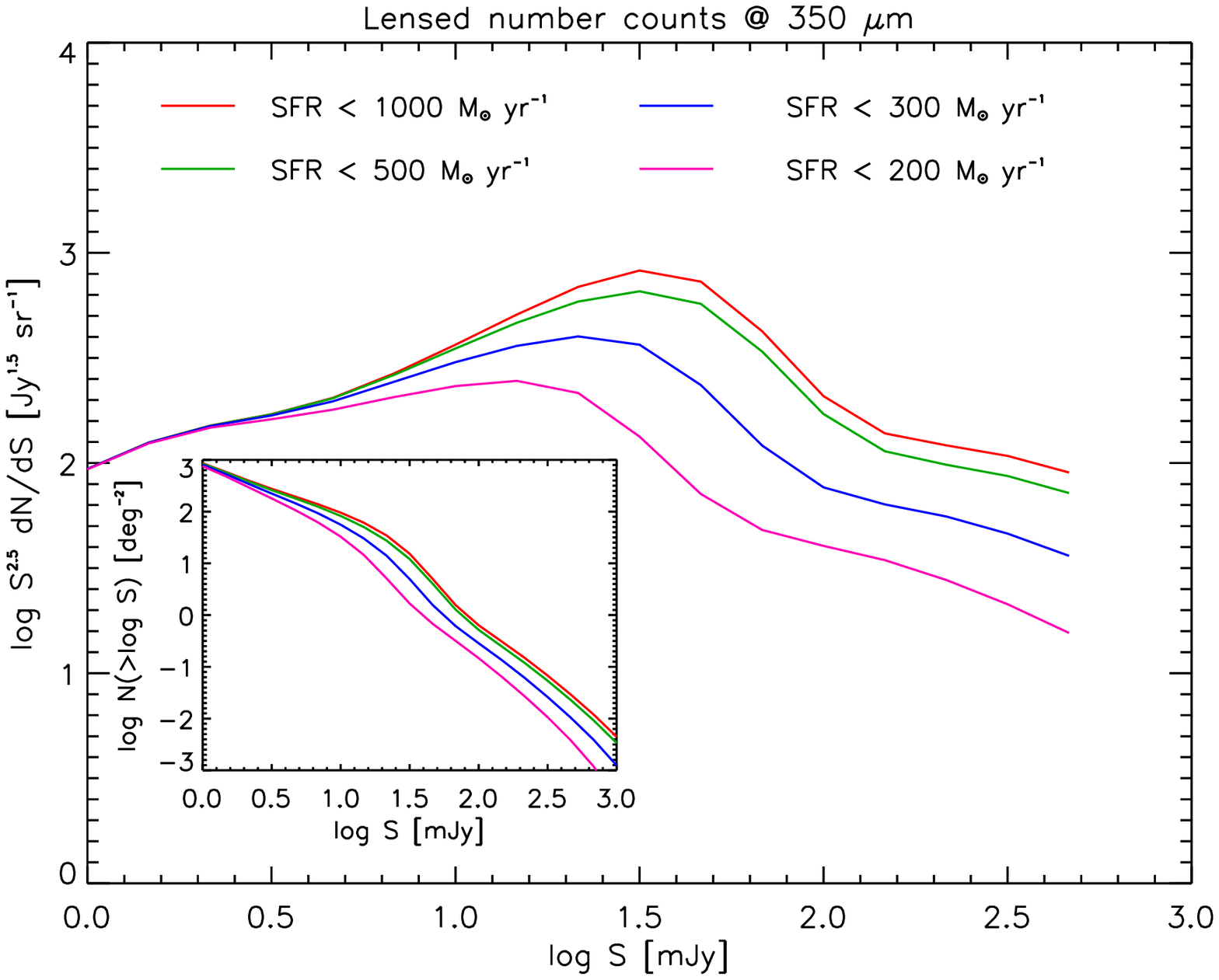}\\
\includegraphics[height=10cm]{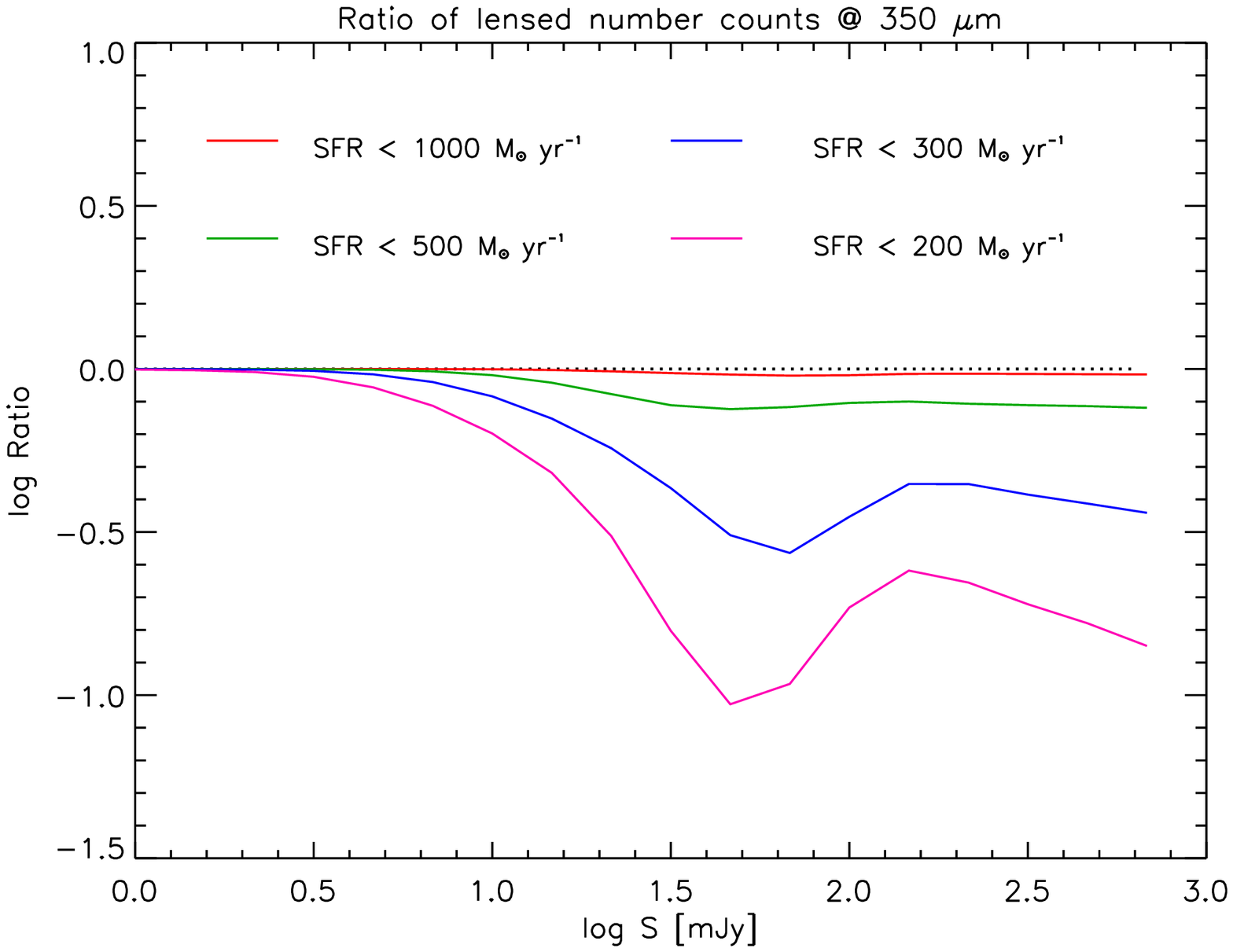}
\caption{\textit{Top panel}: Euclidean normalized counts of lensed galaxies
at $350\,\mu$m. Our model predictions on the counts of lensed
proto-spheroidal galaxies for the SISSA model are plotted for different upper
limits to the star formation rate: SFR $\le  200$, $300$, $500$, and
$1000 \,M_\odot$ yr$^{-1}$ (magenta, blue, green, and red lines,
respectively); the latter line is undistinguishable from that without SFR limit.
In the inset the corresponding integral counts are plotted. \textit{Bottom
panel}: ratio between the counts of lensed galaxies with SFRs below the above
thresholds and those without any SFR limit.}\label{fig:counts350_sfr}
\end{figure}

\begin{figure}
\center
\includegraphics[width=5cm]{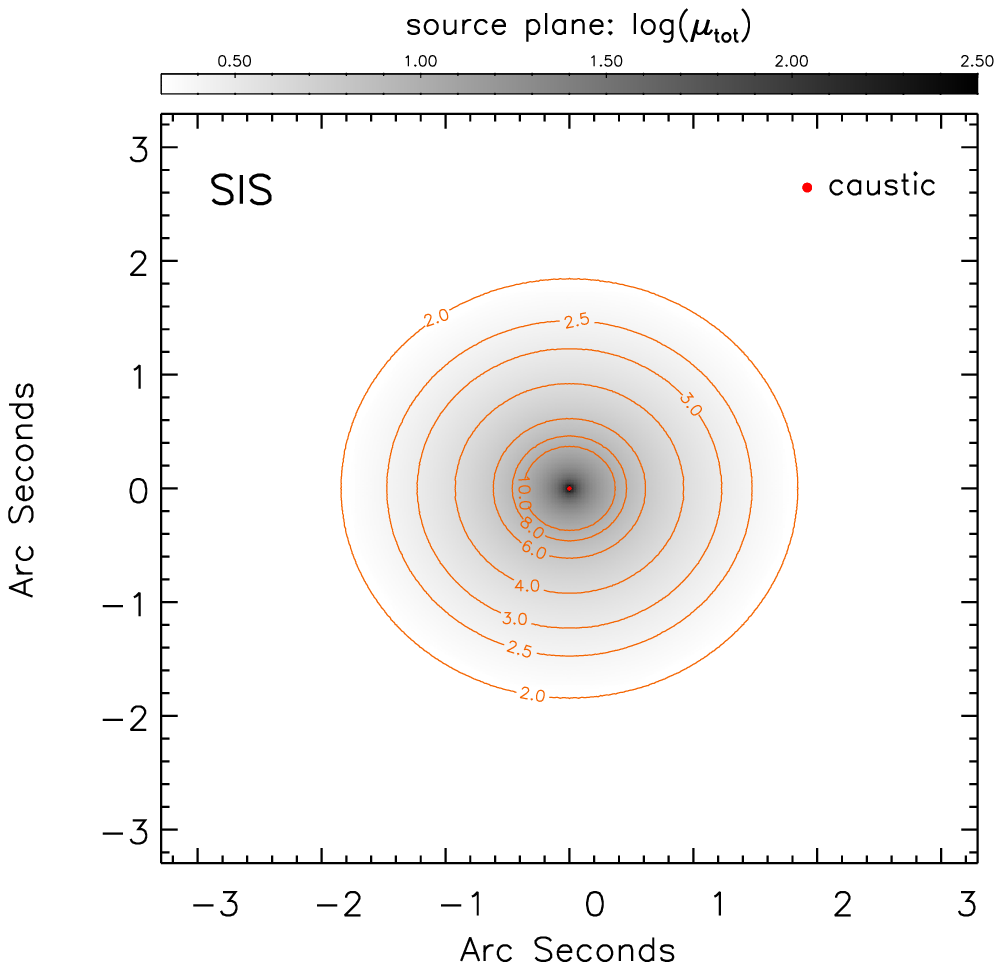}
\includegraphics[width=5cm]{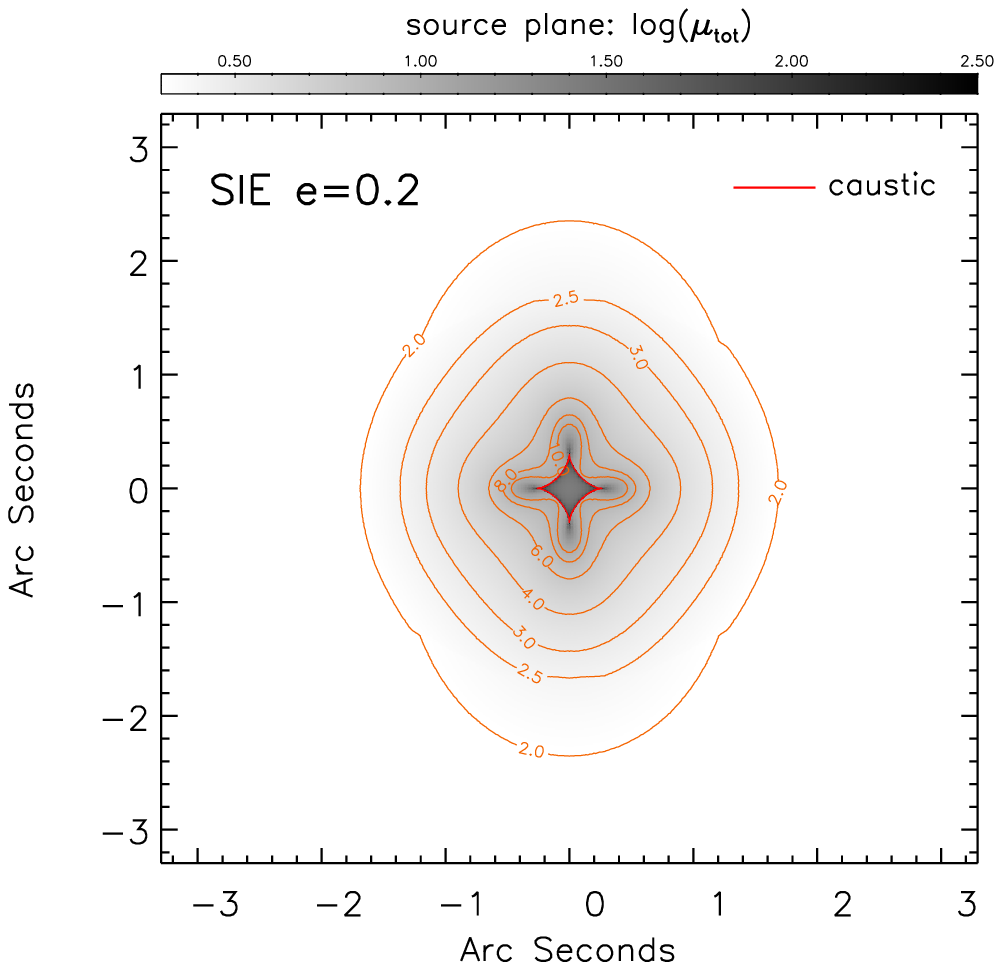}
\includegraphics[width=5cm]{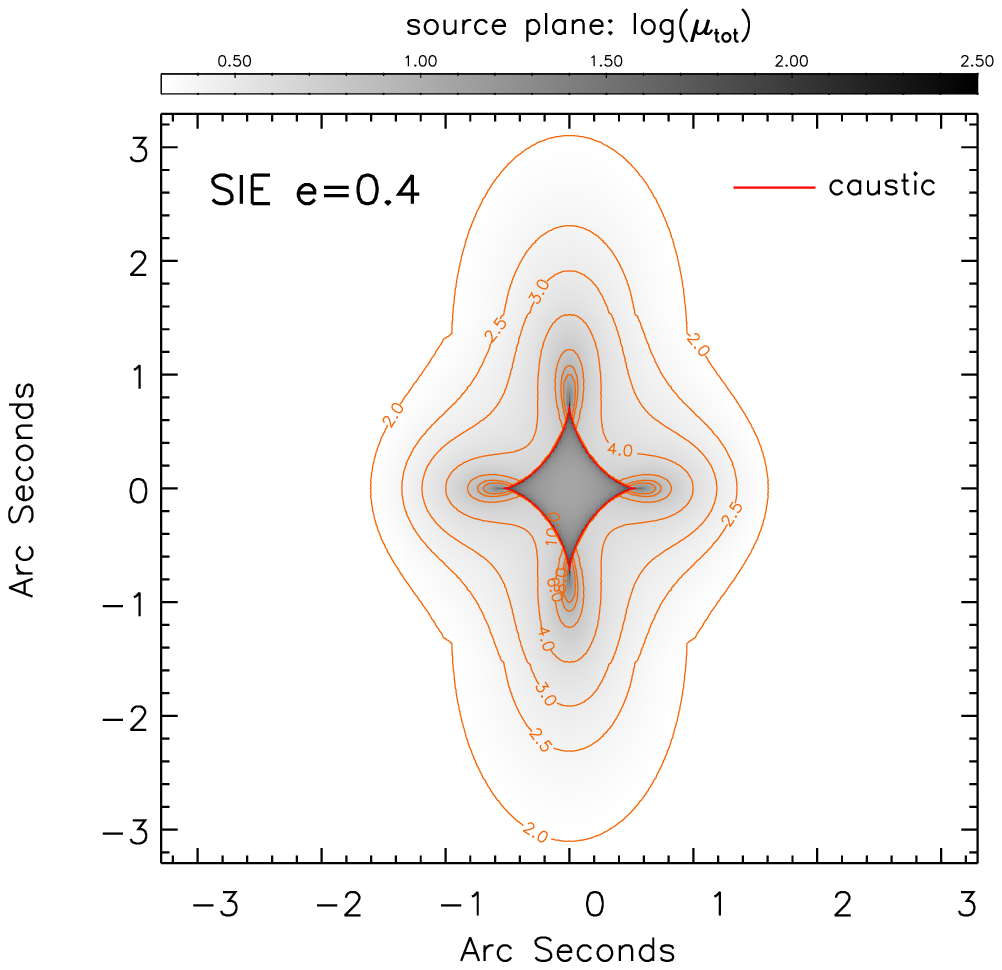}\\
\bigskip\bigskip
\includegraphics[height=10cm]{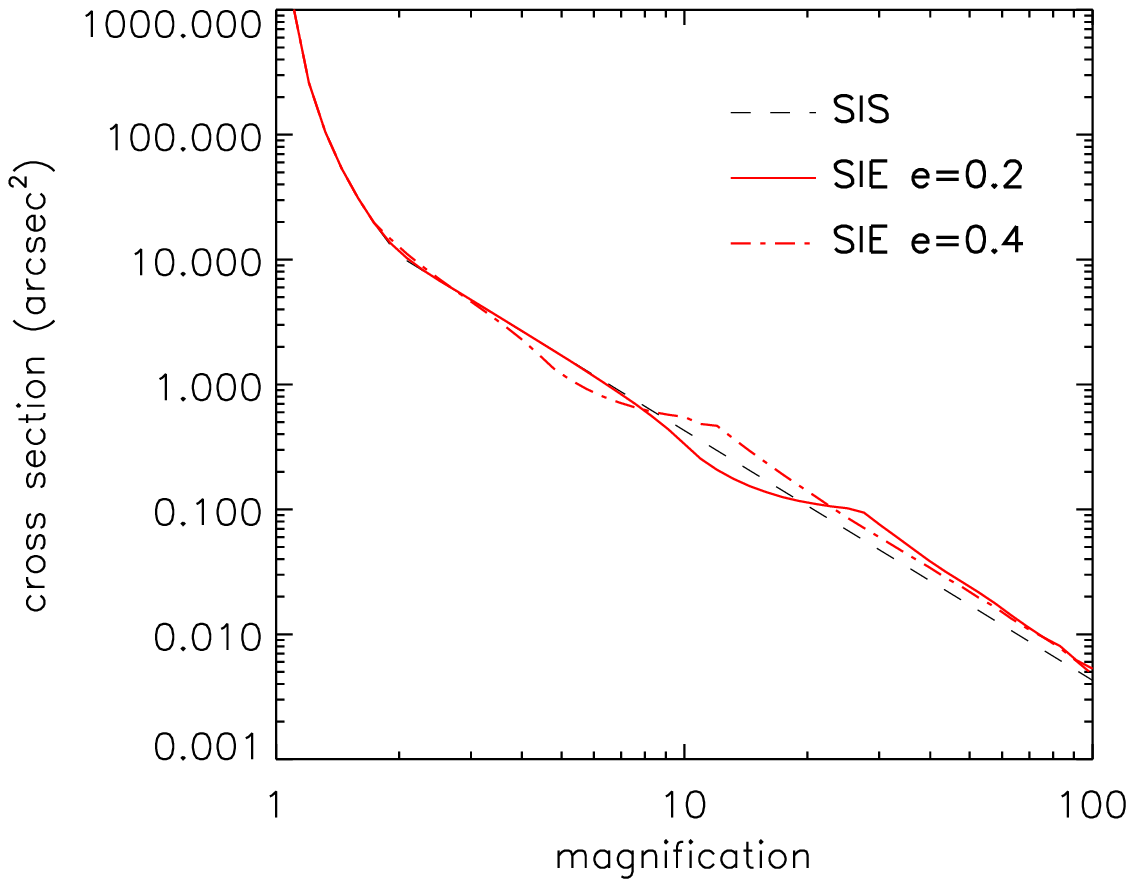}\\
\caption{\textit{Top panels}: map of the total amplification as a function of
the source position from the center of the lensing mass, for a SIS lens
(left), and for a SIE lens with ellipticity $\mathrm{e}=0.2$ (middle)
and $\mathrm{e}=0.4$ (right). The lens and the source are assumed to be at
redshift $0.7$ and $2.5$, respectively. The orange lines represent contours
of equal amplifications ranging from $2$ to $10$, while the red point/curve
mark the caustics. \textit{Bottom panel}: corresponding cross sections for
lensing as a function of the total amplification for the SIS lens (black
dashed line), and for the SIE lens with $\mathrm{e}=0.2$ (red solid
line) and $\mathrm{e}=0.4$ (red dot-dashed line).}\label{fig:ellipticity}
\end{figure}

\begin{figure}
\plotone{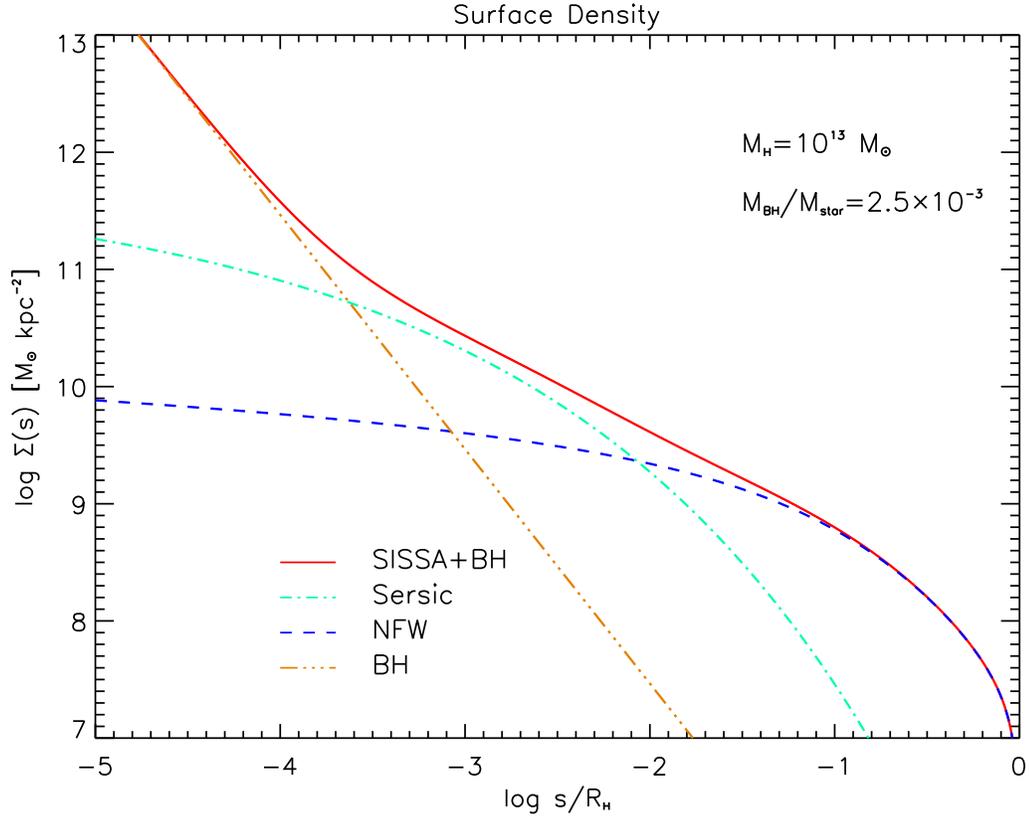} \caption{Surface density profile of a configuration
constituted by an early-type galaxies and a central super-massive BH. This
figure should be contrasted with Fig.~\protect\ref{fig:surface_dens} which
represents the same configuration without central super-massive BH. (Blue)
dashed line: dark matter component with mass $M_{\rm H}=10^{13}\, M_{\odot}$
and NFW profile with concentration parameter $c=5$. (Cyan) dot-dashed line:
stellar component in the proportion $M_{\rm H}/M_\star=30$ relative to the DM
with a S\'ersic $n=4$ profile. (Orange) triple-dot-dashed line: super-massive
BH with mass ratio $M_\bullet/M_\star=2.5\times 10^{-3}$. (Red) solid line:
SISSA$+$BH model, sum of the three contributions. Note that here, to
avoid representing the central singularity, we rendered the point mass
surface density with the powerlaw $\Sigma(\theta)\propto \theta^{-\eta}$ in
the limit $\eta\rightarrow 2$ (see \S~5.4).}\label{fig:BH1}
\end{figure}

\begin{figure}
\plotone{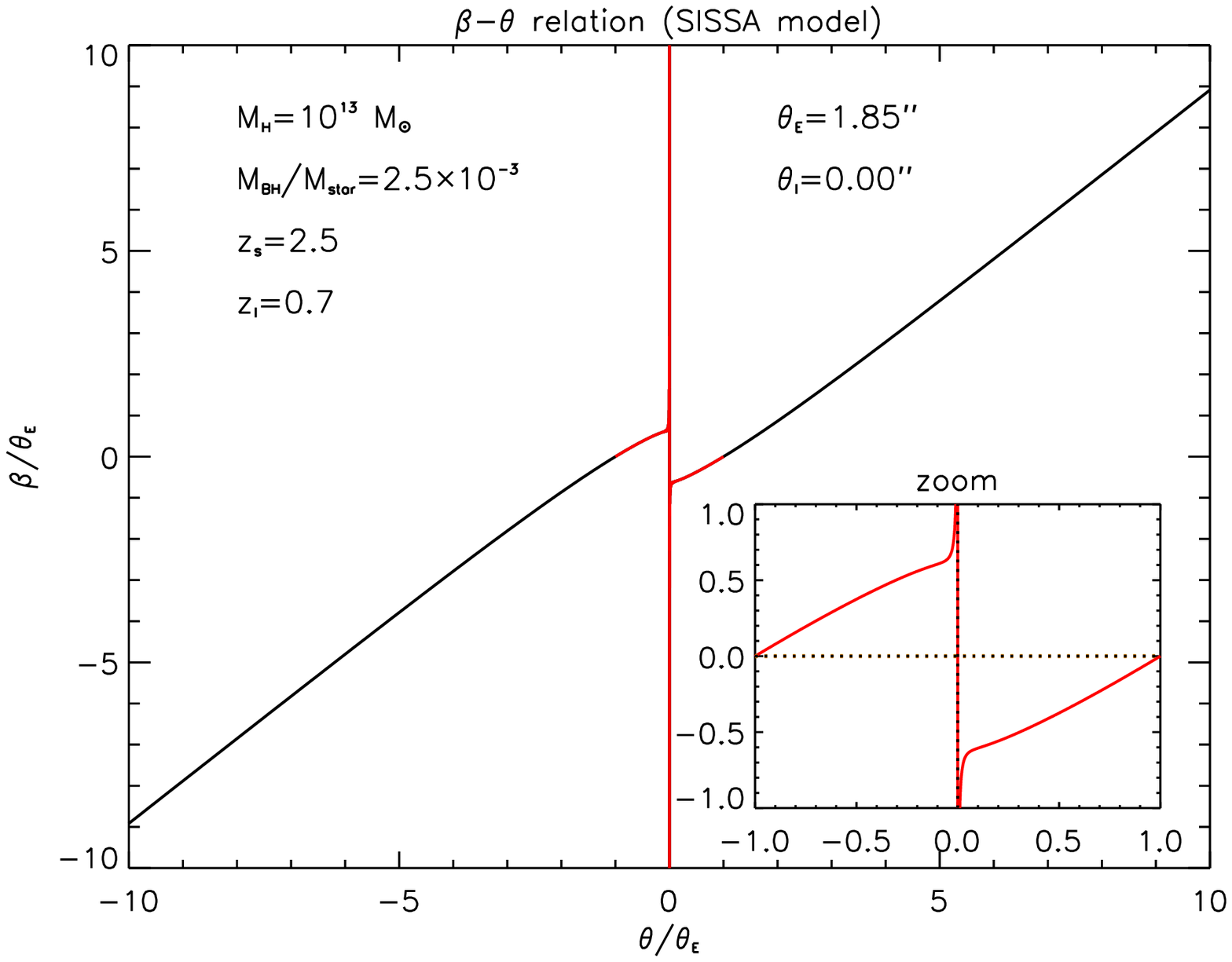}\\
\plotone{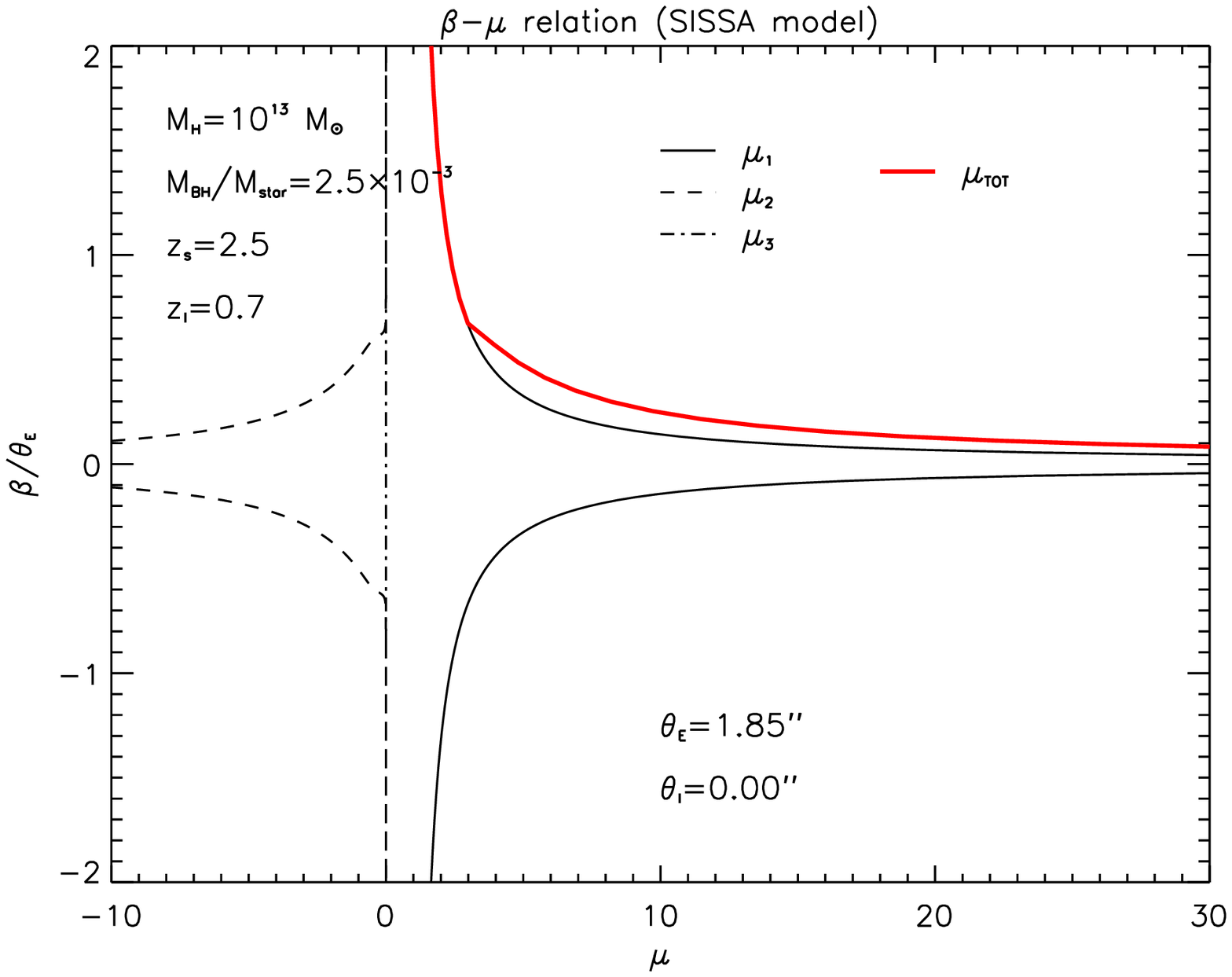}
\caption{Solutions of the lensing equation for the SISSA model, including a
central super-massive BH with mass ratio $M_\bullet/M_\star=2.5\times
10^{-3}$. This figure should be contrasted with
Fig.~\protect\ref{fig:solutions}, which corresponds to the same configuration
without central super-massive BH. Linestyles as in
Fig.~\protect\ref{fig:solutions}.}\label{fig:BH2}
\end{figure}

\begin{figure}
\plotone{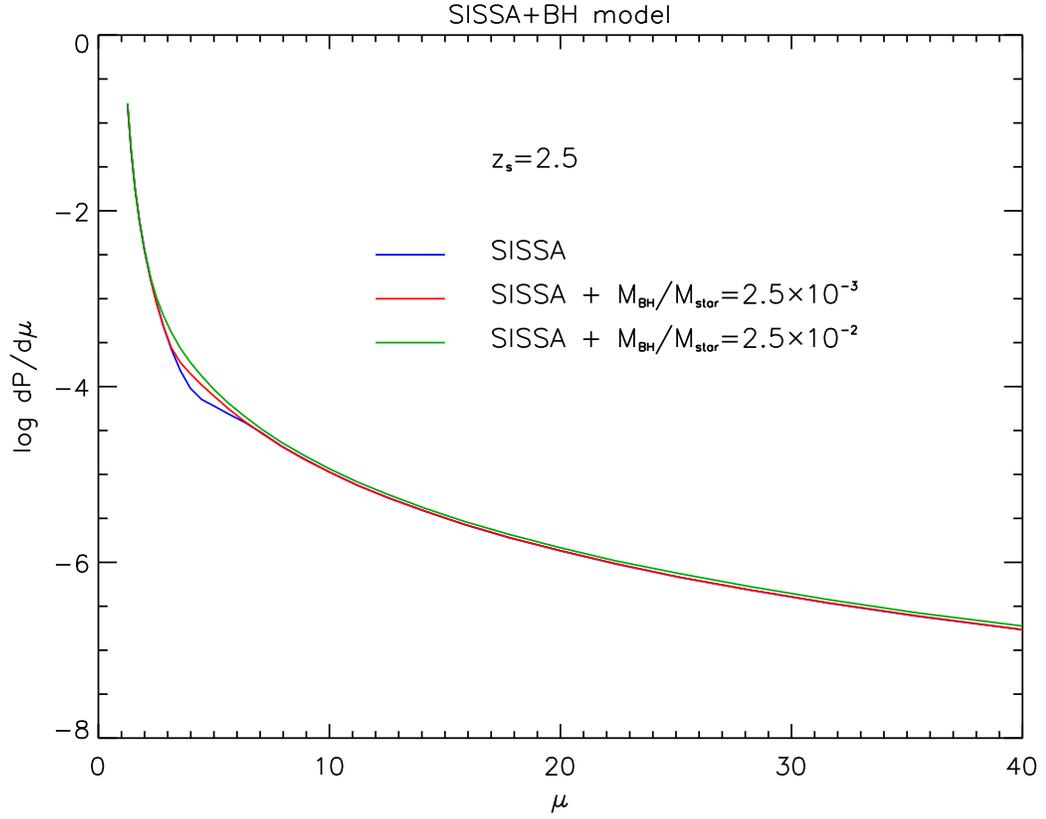} \caption{Differential amplification distribution for a
configuration constituted by an early-type galaxies and a central
super-massive BH. Blue line: SISSA model with fiducial parameter values,
without central super-massive BH. Red line: SISSA model plus a super-massive
BH with standard mass ratio $M_{\bullet}/M_\star=2.5\times 10^{-3}$. Green
line: SISSA model plus a super-massive BH with a high mass ratio
$M_{\bullet}/M_\star=2.5\times 10^{-2}$.}\label{fig:BH3}
\end{figure}

In view of the large samples of strongly lensed galaxies that are being/will
be provided by large area sub-mm (Serjeant 2011; Negrello et al. 2010;
Gonz\'alez-Nuevo et al. 2012), optical (e.g., Oguri \& Marshall 2010) and
radio (SKA) surveys (e.g., Koopmans et al. 2004)  we have worked out simple
analytical formulae that accurately approximate the relationship between the
position of the source with respect to the lens center and the amplification
of the images and hence the cross section for lensing (see
Fig.~\ref{fig:solutions}). The approximate relationships are based on a lens
matter density profile appropriate for early-type galaxies, that comprise
most of the lenses found with different selection criteria. The adopted
profile is a combination of a S\'ersic profile, describing the distribution
of stars, with a NFW profile for the dark matter. We find that, for
essentially the full range of parameters either observationally determined
(for the S\'ersic profile) or yielded by numerical simulations (for the NFW
profile), the combination can be very well described, for lens radii relevant
for strong lensing, by a simple power law. Remarkably, the power law slope is
very weakly dependent on the parameters characterizing the matter
distribution of the lens (the dark matter to stellar mass ratio, the S\'ersic
index, the concentration of NFW profile). For the most common parameter
choices, the slope is slightly sub-isothermal if we consider the projected
profile and slightly super-isothermal if we consider the 3-dimensional
profile, in good agreement with the results of detailed studies of individual
lens galaxies (e.g. Koopmans et al. 2009; Spiniello et al. 2011; Barnab\'e et
al. 2011; Ruff et al. 2011; Grillo 2012; Bolton et al. 2012). Our approach
implies slightly steeper slopes of the total matter density profile for the
least massive systems (see Table~\ref{tab:fits}); evidence in this direction
has been reported by Barnab\'e et al. (2011).

Table~\ref{tab:fits} shows that, if the source and lens redshifts are
measured and the halo mass of the lens is reliably estimated, the factor
$[2\,\Sigma_0/(2-\eta)\,\Sigma_c]^{1/\eta}$, and hence $\theta_E$ [see
Eq.~(\ref{eq:thetaE})], varies by no more than $20-30\%$ for conceivable
variations of the parameters of the lens mass distribution. Such small
variance paves the way to the possibility of exploiting gravitational lensing
as a probe of cosmological parameters (Grillo et al. 2008).

Our simple analytic solutions provide an easy insight into the role of the
different ingredients that determine the lens cross section and the
distribution of gravitational amplifications. The maximum amplification
depends primarily on the source size. Amplifications larger than $\approx
20$, as found for some sub-mm and optical sources (Belokurov et al. 2007;
Negrello et al. 2010; Swinbank et al. 2010; Brownstein et al. 2012), are
indicative of compact source sizes at high-$z$, in agreement with
expectations if most of the stars formed during dissipative collapse of cold
gas. Similarly, analytic formulae highlight in a transparent way the role of
parameters characterizing the lens mass profile ($M_{\rm H}$, $M_{\rm
H}/M_\star$ ratio, concentration of the DM component, S\'ersic index of the
stellar component), and of the source and lens redshifts. They also allow a
fast application of ray-tracing techniques to model the effect of lensing on
a variety of source structures. We have investigated, in particular, the
cases of a point-like or of an extended source with a smooth profile, and of
a source comprising various emitting clumps (as frequently found for high-$z$
active star-forming galaxies). Our formalism has allowed us to reproduce the
counts of strongly lensed galaxies found in the H-ATLAS SDP field.

While our analysis is focussed on spherical lenses, we have also discussed
the case of disk galaxies (showing why they are much less common, even though
late-type galaxies are more numerous) and the effect of ellipticity.
Furthermore we have discussed the effect of a cluster halo surrounding the
early-type lens and of a supermassive BH at its center.

\begin{acknowledgements}
The work has been supported in part by ASI/INAF Agreement I/072/09/0 for the
\textsl{Planck} LFI activity of Phase E2, by INAF through the PRIN 2009 ``New
light on the early Universe with sub-mm spectroscopy'', and by MIUR PRIN
2009. J.G.-N. acknowledges financial support from the Spanish CSIC for a
JAE-DOC fellowship and partial financial support from the Spanish Ministerio
de Ciencia e Innovacion, project AYA2010-21766-C03-01. We thank the
referee for helpful comments and suggestions. We acknowledge useful
discussions with M. Massardi, F. Perrotta, and P. Salucci. A.L. thanks SISSA
for warm hospitality.
\end{acknowledgements}

\clearpage

\clearpage

\clearpage

\clearpage

\clearpage

\clearpage

\clearpage

\clearpage

\clearpage

\clearpage

\clearpage

\clearpage

\clearpage

\clearpage

\clearpage

\clearpage

\clearpage

\clearpage

\clearpage

\clearpage
\begin{deluxetable}{ccccccccccccccccc}
\tabletypesize{}\tablecaption{Power law fits of the surface density profile}
\tablewidth{0pt}\tablehead{\multicolumn{5}{c}{Lens parameter} &&
\multicolumn{3}{c}{$M_{\rm H}=10^{13}\, M_{\odot}$} &&
\multicolumn{3}{c}{$M_{\rm H}=10^{12}\, M_{\odot}$} &&
\multicolumn{3}{c}{$M_{\rm H}=10^{11}\, M_{\odot}$}\\
\cline{1-5}\cline{7-9}\cline{11-13}\cline{15-17}\\
\colhead{$M_{\rm
H}/M_\star$} && \colhead{$n$} && \colhead{$c$} && \colhead{$\log \Sigma_0$}
&& \colhead{$\eta$} && \colhead{$\log \Sigma_0$} && \colhead{$\eta$} &&
\colhead{$\log \Sigma_0$} && \colhead{$\eta$}} \startdata
$30$ && $4$  && $5$  && $9.621$ && $0.823$ &&  $9.326$ && $0.871$ && $9.013$ &&
$0.902$\\
\\
$10$ && $4$  && $5$  && $9.780$ && $0.936$ &&  $9.537$ && $1.036$ && $9.275$ &&
$1.124$\\
$50$ && $4$  && $5$  && $9.561$ && $0.773$ &&  $9.249$ && $0.801$ && $8.922$ &&
$0.814$\\
$70$ && $4$  && $5$  && $9.528$ && $0.743$ &&  $9.208$ && $0.761$ && $8.874$ &&
$0.765$\\
L+11 && $4$  && $5$  && $9.610$ && $0.815$ &&  $9.351$ && $0.893$ && $9.095$ &&
$0.977$\\
\\
$30$ && $3$  && $5$  && $9.629$ && $0.828$ &&  $9.348$ && $0.892$ && $9.042$ &&
$0.934$\\
$30$ && $6$  && $5$  && $9.604$ && $0.811$ &&  $9.295$ && $0.841$ && $8.975$ &&
$0.861$\\
$30$ && $10$ && $5$  && $9.578$ && $0.789$ &&  $9.258$ && $0.806$ && $8.932$ &&
$0.816$\\
\\
$30$ && $4$  && $2$  && $9.472$ && $0.793$ &&  $9.189$ && $0.855$ && $8.882$ &&
$0.895$\\
$30$ && $4$  && $10$ && $9.912$ && $0.982$ &&  $9.601$ && $1.011$ && $9.280$ &&
$1.030$\\
$30$ && $4$  && $c(M)$ && $9.621$ && $0.823$ && $9.369$ && $0.886$ && $9.101$ &&
$0.932$\\

\enddata
\tablecomments{Power law fits of the surface density profile
[Eq.~(\protect\ref{eq:power_law}) over the radial range $10^{-2.5}\la
s/R_{\rm H}\la 10^{-1}$ for halo masses $M_{\rm H}=10^{11}$, $10^{12}$, and
$10^{13}\, M_{\odot}$. The normalization $\Sigma_0$ (in $M_\odot$ kpc$^{-2}$)
refers to a projected radius $s_0=10^{-2}\, R_{\rm H}$. The table shows the
dependence of the fits on the parameters of the lens mass distribution,
namely, dark matter to star mass ratio $M_{\rm H}/M_\star$, S\'ersic index
$n$, and halo concentration $c$. The case L+11 takes into account the
dependence of ratio $M_{\rm H}/M_\star$ on $M_{\rm H}$ as implied by the Lapi
et al. (2011) model.}\label{tab:fits}
\end{deluxetable}

\end{document}